\documentclass[letterpaper,twocolumn,10pt]{article}
\usepackage{custom}

\usepackage{amsmath, amssymb, amsthm, bm}
\usepackage{graphicx}
\usepackage[dvipsnames]{xcolor}
\graphicspath{ {images/} }
\usepackage{caption}
\usepackage{subcaption}
\usepackage{multirow}
\usepackage{enumitem}
\usepackage{siunitx}
\usepackage{algorithm}
\usepackage[noend]{algorithmic}
\usepackage{mathtools}
\usepackage{color, url}
\usepackage{calrsfs}
\usepackage[font=normalsize, labelfont={bf}, skip=3pt]{caption}
\usepackage{xurl}
\usepackage{xspace}

\usepackage[usenames, dvipsnames]{pstricks}
\usepackage{pstricks-add}
\usepackage{epsfig}
\usepackage{pst-grad} 
\usepackage{pst-plot} 
\usepackage[space]{grffile} 
\usepackage{etoolbox} 
\usepackage{booktabs} 
\usepackage{threeparttable} 
\makeatletter 
\patchcmd\Gread@eps{\@inputcheck#1 }{\@inputcheck"#1"\relax}{}{}
\makeatother

\renewcommand{\v}{{\mathsf{val}}}

\newcommand{\tsc}{T_{\sf{S}}^{\sf{coin}}}
\newcommand{\tac}{T_{\sf{A}}^{\sf{coin}}}
\newcommand{\tsg}{T_{\sf{S}}^{\sf{gen}}}
\newcommand{\tagg}{T_{\sf{A}}^{\sf{gen}}}
\newcommand{\tsf}{T_{\sf{S}}^{\sf{fake}}}
\newcommand{\taf}{T_{\sf{A}}^{\sf{fake}}}
\newcommand{\ta}{T_{\sf{A}}}

\newcommand{\tpa}{T_{\sf{A}}(p)}

\newcommand{\sgl}{{\sf{GL}}}
\newcommand{\smfpg}{{\sf{MFPG}}}
\newcommand{\smppg}{{\sf{MFPPG}}}

\newcommand{\scc}{{\sf{CC}}}
\newcommand{\scg}{{\sf{CG}}}
\newcommand{\spub}{{\sf{PS}}}
\newcommand{\sfr}{{\sf{FR}}}

\newcommand{\sft}[1]{{\sf{#1}}}

\newcommand{\ph}{\sft{Phishing}}
\newcommand{\tg}{\sft{Target}}
\newcommand{\bn}{\sft{Benign}}

\newcommand{\val}{\sft{val}}

\newcommand{\don}{\sft{d}_1}
\newcommand{\dt}{\sft{d}_2}

\newcommand{\w}{\sft{w}}

\newcommand{\lf}{\sft{lf}}

\newcommand{\mo}{\sft{m}_1}
\newcommand{\mt}{\sft{m}_2}
\newcommand{\po}{\sft{p}_1}
\newcommand{\pt}{\sft{p}_2}

\newcommand{\sct}[1]{{\sc{#1}}}
\newcommand{\flf}{\sct{FindLaunderingFrom}}
\newcommand{\dlf}{\sct{DynamicLaunderingFlow}}

\newcommand{\coh}{\sft{coh}}

\newcommand{\Li}{L_{\text{in}}}
\newcommand{\Lo}{L_{\text{out}}}
\newcommand{\LTC}{L_{\text{TC}}}

\newcommand{\lo}{\ell_{\text{out}}}
\newcommand{\To}{T_{\text{out}}}
\newcommand{\so}{\sigma_{\text{out}}}
\renewcommand{\si}{\sigma_{\text{in}}}

\newcommand{\ke}{k_{\textsf{eff}}}
\newcommand{\csh}{C_{\textsf{SH}}}
\newcommand{\cis}{C_{\textsf{IS}}}

\newcommand{\la}{\leftarrow}

\theoremstyle{definition}
\newtheorem{example}{Example}
\newtheorem{definition}{Definition}

\newcommand{\rev}[1]{{\color{black}{#1}}}

\renewcommand{\red}[1]{\color{black}({#1})}
\newcommand{\algcomment}[1]{\textcolor{gray}{\scriptsize{\#~#1}}}

\newcommand{\et}{{\it et al.}}

\newcommand{\txlink}[1]{%
    \footnote{\href{https://etherscan.io/tx/#1}{\scriptsize\path{tx:#1}}}%
}

\newcommand{\txlinkLine}[1]{%
    {\href{https://etherscan.io/tx/#1}{\path{#1}}}%
}

\newcommand{\alink}[1]{%
    \footnote{\href{https://etherscan.io/address/#1}{\scriptsize\path{#1}}}%
}
\newcommand{\alinkLine}[1]{%
    {\href{https://etherscan.io/address/#1}{\path{#1}}}%
}

\newcommand{\ETH}{\texttt{ETH}\xspace}
\newcommand{\WETH}{\texttt{WETH}\xspace}

\newcommand{\stETH}{\texttt{stETH}\xspace}
\newcommand{\USDT}{\texttt{USDT}\xspace}
\newcommand{\USDC}{\texttt{USDC}\xspace}

\newcommand{\DAI}{\texttt{DAI}\xspace}

\newcommand{\WBTC}{\texttt{WBTC}\xspace}

\newcommand{\BUSD}{\texttt{BUSD}\xspace}

\begin{document}

\date{}

\title{\Large \bf The Anatomy of Address Poisoning on Ethereum:\\ Funding Mechanisms, Scam Signatures, and  Laundering via Tornado Cash}

\author{
{\rm Son Hoang Dau}\\
RMIT University
\and
{\rm Thanh Nguyen}\\
RMIT University
\and
{\rm Phuong Duy Huynh}\\
RMIT University
\and 
{\rm Hong Yen Tran}\\
CSIRO
\and
{\rm Nicholas Huppert}\\
RMIT University
\and
{\rm Jeff Nijsse}\\
RMIT University
\and
{\rm Huong Ha}\\
RMIT University
\and
{\rm Xun Yi}\\
RMIT University
} 

\maketitle
   
    \begin{abstract}
        Address-Poisoning Transfer (APT) is a prevalent blockchain phishing scam in which a scammer poisons a victim's address book by generating a transfer with a phishing address that looks similar to a benign address that the victim has previously interacted with. 
        Although simple, APT phishing attacks have cost users millions of dollars in recent years, which has captured the attention of the research community (Ye \textit{et al.} WWW'24, Guan-Li CCS'24, Chen \textit{et al.} NDSS'25, Tsuchiya \textit{et al.} USENIX'25). 
        In this work, we go beyond detection and investigate three important and underexplored aspects of APT: scam funding mechanisms, scam signatures, and scam proceeds laundering via public services.  
        In particular, we propose five families of scam signatures that capture key aspects of APT operations, which are useful for address clustering.
        \rev{We also conduct the first investigation into usage of Tornado Cash for funding APTs and laundering scam proceeds.}

    \end{abstract}

    \section{Introduction}


    Global crypto scam revenue surpassed \$10 billion in 2021 and has continued to rise steadily according to the Chainalysis 2025 Crypto Crime Report~\cite{Chainalysis_crypto_crime_report_2025}. Ranked fifth among all scams in 2024 revenue by Chainalysis, Address-Poisoning Transfer (APT) is a widespread scam (see, e.g.~\cite{Binance_20M_phishing, Chainalysis_address_poisoning_2024}) that exploits a combination of human's limitations in remembering long address strings, their copy-and-paste habit, and the fact that most digital wallets only display the first and last few digits of their addresses. Also referred to as dust-value transfer (DVT), in its simplest form the scammer generates a \textit{phishing} address $p$ that looks similar to a \textit{benign} address $b$ that the victim address $v$ has previously interacted with and transfers a tiny amount of a high-value token from $p$ to $v$, hoping that the victim would later mistakenly transfer funds to $p$ instead of the intended address. For example, a notable DVT scam occurred in May 2024 in which a crypto whale mistakenly transferred 1,155 \WBTC (about \$70 million) to a phishing address, which has the first four and the last five digits identical to the victim's intended address~\cite{Chainalysis_address_poisoning_2024}. In another major incident, 
    the victim lost 2,000 \ETH (about \$4.4 million) after receiving several fake-token transfers (see Example~\ref{ex:2000ETH}). 

    \begin{figure*}
        \includegraphics[width=1\textwidth]{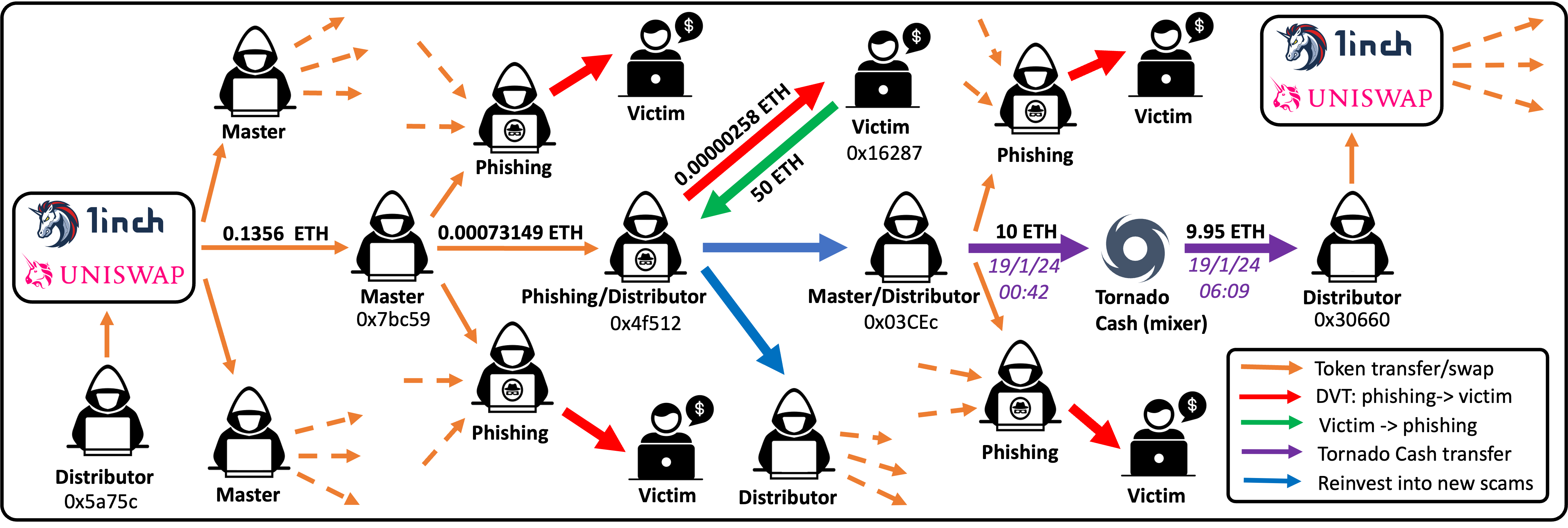}
        \caption{An APT scam cluster with distinctive patterns: distributor-master-phishing funding hierarchy, distributors funding masters via 1inch or \ETH transfers, APT amounts consistently in $[10^{-6},10^{-5})$. Notably, the phishing address \protect\alinkLine{0x4f512487a746AB1638B5fFb0D8321dBDFA6Fb8eA}, after receiving 50 \ETH from a victim, reinvested by funding other distributors/masters. The master  \protect\alinkLine{0x03CEc7583846EEEe04cDd0752A0B35bC8F793bff}, besides funding phishing addresses, sent 10 \ETH via Tornado Cash to a distributor \protect\alinkLine{0x306608Bf57e58A1263c491aD81b1A5fA2B420cfd}, which then started a new cluster with an identical pattern.}
        \label{fig:example_DVT_network_TC}
    \end{figure*}

%

    Existing work on APT scams focuses primarily on scam detection and victim analysis~\cite{YeHongZhangYang_WWW24, GuanLi_CCS24, Chen_etal_NDSS2025, tsuchiya_USENIX_2025}. This leaves several critical questions insufficiently explored, such as: What behavioural patterns are exhibited by scam addresses? How does scam funding operate? \rev{And to what extent are scam funding and proceed laundering carried out via public services like Tornado Cash?}
    To address these gaps, we take a step toward a more comprehensive understanding of APT scams by examining three key aspects: \textit{scam funding mechanisms}, \textit{scam signatures}, and \textit{scam proceeds laundering}. 
    
    For \textit{scam funding mechanisms}, we first establish a comprehensive classification of eleven types of APTs (see Fig.~\ref{fig:APT_subtypes_masters}), instead of three types as in the literature. Then, based on this finer classification, we describe the funding mechanism for each APT type, noting that different APT types require different funding mechanisms  (Sections~\ref{sec:APT_datasets}, \ref{sec:master_distributor_datasets}). For instance, while a phishing address deploying coin APTs (where APTs are in \ETH) consistently requires a coin funder, an address deploying genuine-token APTs may not. Coin funding is only necessary if the token amount is nonzero, requiring \ETH to approve a contract to spend tokens on its behalf, or if the address initiates the APT itself, requiring \ETH to cover transaction fees.

    Equipped with new insights into APT scams, particularly their funding and attack mechanisms, we introduce five sets of \textit{scam signatures} 
    tailored to APT scams (Section~\ref{sec:signature_extraction}). These signatures capture: (i) the contracts created and invoked by scam addresses, (ii) APT-value patterns, (iii) the relationships among funding values, APT values, and transaction fees, (iv) gas configurations used in APT-carrying transactions, and (v) the use of public services by scam addresses.

    Scam signatures are important because they could potentially provide evidence that seemingly distinct scam activities are likely operated by the same scammer group. 
    For example, in the scam cluster shown in Fig.~\ref{fig:example_DVT_network_TC}, beyond the explicit linkage via \ETH transfers (an indicator largely overlooked in the APT literature), we observe consistent behavioural patterns: these addresses repeatedly use 1inch to route funds to the funders of phishing addresses, and the APT amounts consistently contain exactly five zeroes after the decimal point.
    \rev{Such signatures could also be useful in tracing through mixers like Tornado Cash when these behavioural patterns persist.} 
    
    \rev{In Section~\ref{sec:TC_usage}, we present an attempt to group Tornado Cash users associated with scam activities, by introducing APT-specific heuristics that can establish connections among otherwise unrelated addresses. We note that \textit{address grouping/clustering} is a fundamental problem in blockchain intelligence research (see, e.g. \cite{HarriganFretter_2016,MoserNarayanan_FC_2022,LinoyStakhanovaMatyukhina_CNSM_2019,Victor_FC_2020,Beres_etal_DAPPS_2021}). We also implement a tracing algorithm  to estimate the total APT proceeds flowing into public services. The algorithm can keep track of individual tokens during laundering and handle token swaps.}

    Our main contributions are summarised below. 
    \begin{itemize}[leftmargin=*]
        \item We propose a classification of 11 sub-types of APT scams and an in-depth study of their funding/attack mechanisms.
        \item \rev{We introduce five families of scam signatures capturing key operational aspects of APTs, and show that they are consistent, i.e. individual addresses tend to reuse the same signature values, suggesting automated scam operation.}
        \item \rev{We conduct the first investigation into how APT scammers utilise Tornado Cash (among public services) for funding and laundering. We also study how Tornado Cash user addresses can be grouped via APT operational connections. In particular, we provide the first estimate of roughly \$38.09M APT scam proceeds entering Tornado Cash and \$18.14M to other public services (11/2022--5/2025).}        
    \end{itemize}

\section{Background and Data Collection}
\label{sec:data_collection}

    \textbf{Ethereum.} Launched in 2015, Ethereum is a decentralised blockchain that enables the creation of smart contracts that are capable of executing business logic~\cite{Buterin2013}. The ERC-20 standard defines an interface for smart contracts to issue fungible tokens, ensuring interoperability between decentralised applications and various virtual assets, including shares, stablecoins, and user generated tokens. Each ERC-20 token has a name, e.g. \texttt{Tether USDT}, a symbol, e.g. \USDT, and a contract address of its smart contract.
    
    Ethereum assets can be transferred in two main ways: by transferring the native currency ether (\ETH), referred to as \textit{coin transfers}, or by transferring ERC-20 tokens. One can also swap one asset for another via a Centralised Exchange (CEX) 
    or a decentralised Exchange (DEX). 
    Every transaction requires a digital signature associated with a unique Ethereum address through the user's wallet. All transactions and associated wallet balances are publicly viewable on-chain, making it possible to trace transactions and limit privacy. Tornado Cash~\cite{TC2019} was developed as a decentralised privacy protocol that allows users to unlink the origin and destination of their transactions, by employing zero-knowledge proof~\cite{Groth2016}.

    \textbf{Data Collection.}
    We used the APIs of Etherscan~\cite{Etherscan} and Google BigQuery~\cite{GBQ}
    to fetch Ethereum data from 1/11/2022 to 07/05/2025 corresponding to blocks 15875000--22431083. Note that APTs were first observed about November 2022 (see, e.g.~\cite[Sec.~5]{GuanLi_CCS24} and \cite{binance_square_APT_post_metamask_alert}). We also collected a list of nearly 40 thousand \textit{public service} addresses based on the public labels from Etherscan and manual inspections.

    \begin{figure*}
        \centering
        \includegraphics[scale=0.695]{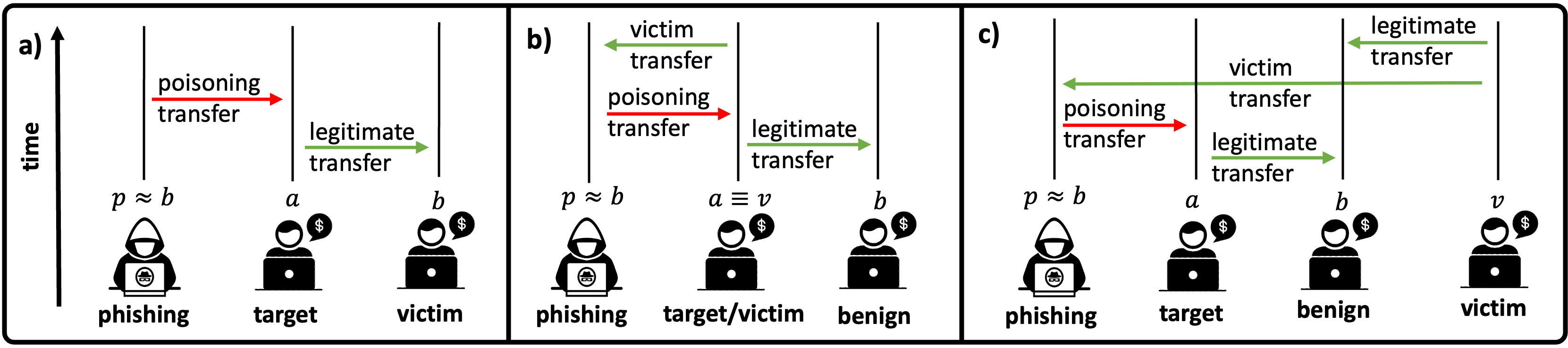}
        \caption{a) The typical APT attack setup: observing a legitimate transfer from $a$ to $b$, the phishing address $p$, which looks similar to $b$, sends an APT to $a$. Note that the APT can also be made from $a$ to $p$ for genuine tokens (must be a zero value) and fake tokens. b) (Type-1) The standard scenario when $a$ mistakenly transfers funds to $p$ instead of the intended receiver $b$. c) (Type-2) A rarer scenario when the victim uses another address $v$ instead of $a$ to (mistakenly) transfer funds to $p$.}
        \label{fig:DVT_victim}
    \end{figure*}
    
    \textbf{Normal and Internal Transactions.} A normal transaction is initiated by an externally owned account (EOA) and is stored directly on-chain as a transaction within a block. In contrast, an internal transaction originates from a contract account (CA) as part of a smart contract execution. Unlike normal transactions, internal transactions are not recorded directly on the blockchain, but can be obtained in the EVM execution trace. Etherscan provides the API to get these transactions by specifying a block range. 
    We collected 1,399,654,309 normal transactions and 639,829,207 internal transactions, occupying 1.7 TB in our local SQL database.

    \textbf{ECR-20 Token Transfers.} A token transfer refers to the movement of a digital token that follows the ERC-20 standard. 
    When a user transfers an ERC-20 token, its smart contract emits a \texttt{transfer} event. We collect all transfer events of the top-100 tokens (ranked by market cap) from CoinGecko~\cite{CoinGecko} including popular ones such as \USDT, \USDC, \DAI, \WETH, \WBTC. We refer to these as \textit{genuine} tokens, as opposed to the \textit{fake} ones that mimic them (see below). Event logs for token contracts are retrieved via the \texttt{getLogs} method, filtered by the topic corresponding to the \texttt{Transfer} event signature, 
    and decoded to extract the sender, recipient, and transfer amount.

    \textbf{Fake Token Transfers.} We adopt the approach by Guan and Li~\cite{GuanLi_CCS24} that assumes that fake tokens often have token symbols resembling those of genuine tokens, while having malicious contracts. We extend and modify their approach to include fake tokens corresponding to the aforementioned top-100 genuine tokens. We generate a list of tokens with unique contract addresses from the publicly available token transfer data set on Google BigQuery. Then, every token that has the same symbol as a genuine token but with a different address is labelled as a fake token. A token is also classified as fake if its symbol is a Levenshtein edit distance of one from a genuine symbol and has either an unverified contract (on Etherscan) or a flawed contract that allows transfers from a zero-balance testing address (see~\cite[Sec.~4.2]{GuanLi_CCS24}). We have collected 55,064 fake tokens. The largest groups mimic \USDT (13,196 - 24\%), PEPE (7,452- 13.5\%), \USDC (6,106 - 11.1\%), \DAI (4,645 - 8.4\%). The total ERC-20 token transfer data occupies 726 GB in our SQL database (see Appendix~\ref{app:data_collection}).

    \textbf{Tornado Cash Transactions.} In Tornado Cash (TC), a user deposits a fixed amount of the available currency (e.g. 0.1, 1, 10, or 100 \ETH) 
    to a TC router contract. The router then forwards the deposit to a corresponding TC pool. This deposit is recorded with other deposits of the same amount, creating an anonymity set. When a user wants to withdraw, they provide their previously generated zero-knowledge proof that proves the ownership of their deposit without revealing their depositing address and request that the funds be withdrawn to a specified address. Optionally, users can withdraw via a third-party service (relayer) to enhance privacy. To collect deposit and withdrawal transactions, we first retrieve all TC pool and router addresses from Etherscan. Then we collect sender and receiver addresses from deposit and withdrawal transactions by decoding their calling functions. We have collected 36,435 deposit and 35,076 withdrawal transactions in TC pools for the period 11/2022--05/2025. \ETH pools are the most active, accounting for about 98\% of the transactions on TC (see also Table~\ref{tab:TC}, Appendix~\ref{app:data_collection}).

    \section{APT Phishing and Victim Datasets}
    \label{sec:APT_datasets}

\begin{figure*}
        \centering
        \includegraphics[width=1.0\textwidth,trim={3.88cm 7.9cm 7.28cm 7.35cm},clip]{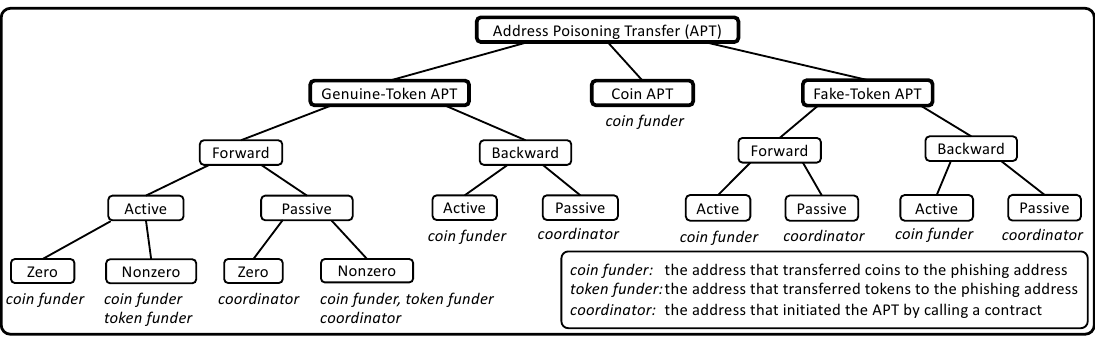}
        \caption{Eleven APT subtypes, classified based on whether the transfer is a coin, genuine-token, or fake-token transfer, whether it is forward (phishing to target) or backward (target to victim), whether it is active (initiated by the phishing address) or passive (initiated by another address), and whether its value is zero. Each APT, depending on the subtype, is supported by either a \textit{coin funder} and/or a \textit{token funder} and/or a \textit{coordinator}, all referred to as \textit{master} addresses.}
        \label{fig:APT_subtypes_masters}
    \end{figure*}

    Prior work on APTs~\cite{YeHongZhangYang_WWW24, GuanLi_CCS24, Chen_etal_NDSS2025, tsuchiya_USENIX_2025} considered three types of APT depending on whether \ETH, genuine ERC-20 tokens, or fake tokens are transferred. We refer to these as \textit{coin} APTs, \textit{genuine-token} APTs, and \textit{fake-token} APTs, respectively. We eventually classify APTs into 11 subtypes (see Section~\ref{subsec:APT_datasets} and Fig.~\ref{fig:APT_subtypes_masters}), which allows us to describe more aspects of APTs. Between 11/2022 and 05/2025 we build an APT dataset consisting of: 3,855,694 coin APTs, 14,799,949 genuine token APTs (using the top-100 ERC-20 tokens from CoinGecko), 37,708,070 fake token APTs (using the 55,064 fake tokens mimicking the 100 genuine tokens). 
    We also build a victim dataset that consists of 4,209 victim transfers, totalling around \$166.31M million USD.
    
    We first describe standard APT scenarios with examples (see also Fig.~\ref{fig:DVT_victim}) before detailing the dataset construction.

    \begin{example}[Type-1 victim]
        \label{ex:68_0_DVT} 
        The phishing address \alinkLine{0xd9A1C3788D81257612E2581A6ea0aDa244853a91}  
        transferred 0 \ETH (\txlinkLine{0x87c6e5d56fea35315ba283de8b6422ad390b6b9d8d399d9b93a9051a3e11bf73}) and also 0.05 fake token 
    (\txlinkLine{0x9147d74ef5749b7f27eb2e2528e5a611060b3f609b435f7f50ac87f49e5b957c}) to the victim \alinkLine{0x1e227979f0b5bc691a70deaed2e0f39a6f538fd5}, mimicking the victim's intended receiver 0xd9A1b...B2853a91, 
    which received 0.05 \ETH from the victim (possibly a test transfer) a few minutes earlier. Due to the high similarity between the two addresses (the first four and last six digits are identical), the victim mistakenly transferred 1,155 \WBTC (about \$70 million USD) to the phishing address about an hour later.

    \end{example}

\begin{example}[Nonzero genuine-token APT]
        \label{ex:multi_DVTs_token} 
        Within a single contract call \txlinkLine{0x057054eb96d386dfbefef7fae2979b3b0e4dd7f742647b47e80c4d5f30974de5} by the labelled address Fake\_Phishing11701 to the contract Fake\_Phishing11700, small amounts of \USDT (e.g. 0.03 and 0.05) were transferred to several phishing addresses and then (the same amounts) to target addresses. 
    \end{example}

\begin{example}[Type-2 victim]
    \label{ex:victim_versus_attacked}
    The phishing address 
   \alinkLine{0x5b9373ba974f0f2b0c99e79004c517fc636ed866} did not target the victim 
   \alinkLine{0xCe7c8d990a248B6E1b9750788a9219497Fa41297}, who still sent the phisher 28.99 \ETH. Prior to this, it attacked another address 
   \alinkLine{0xAeeF025bB585d47D9A1886d1E8795Cba0816B0B6}, and both the victim and the target transferred to a common benign address 
   \alinkLine{0x5B93732311eb61642058fa42202bC4fc496eD866}, which shares the same first and last six characters with the phishing address. 
\end{example}

    \subsection{Detecting the Address Poisoning Transfers}
    \label{subsec:APT_datasets}
 
     Our detection approach shares similarities with Guan-Li~\cite{GuanLi_CCS24},  Chen \et~\cite{Chen_etal_NDSS2025}, and Tsuchiya \et~\cite{tsuchiya_USENIX_2025}, however, captures more nuance by dividing APTs into 11 subtypes shown in Fig.~\ref{fig:APT_subtypes_masters}, providing a more complete understanding of how they work. In particular, having refined subtypes allows for 
     extraction of APT scam signatures, which was not possible with existing approaches (see more in Section~\ref{sec:signature_extraction}).

    We distinguish \textit{active} APTs, in which phishing addresses initiated the transfers (and paid the fees of the transactions containing the APTs), from \textit{passive} APTs, in which coordinator addresses rather than the phishing's initiated the transfers. We also distinguish \textit{forward} APTs, in which the transfers were from the phishing addresses to the target addresses, from \textit{backward} APTs, in which the transfers were from the target addresses to the phishing addresses. We also consider if the value of the APT is zero or nonzero. We note that nonzero-value passive genuine-token APTs require phishing addresses to send \texttt{approve} transactions that allow some contracts to spend the genuine tokens on their behalf, while zero-valued passive genuine-token APTs do not. Existing work overlooks these nuances, which could lead to miscalculations of the costs of different subtypes of APT.        
    This finer classification also allows for more accurate tracing of the addresses directly supporting the phishing addresses. For instance, while a \textit{forward, active, positive-valued, genuine-token} APT has an accompanying \textit{coin funder} and a \textit{token funder}, a \textit{backward, passive, genuine-token} APT only has one accompanying \textit{coordinator} and no funder. This distinction becomes critical when we construct the datasets of funders and coordinators from the datasets of APTs. 
    
    We divide the construction process into three sub-procedures that identify coin APTs, genuine-token APTs, and fake-token APTs separately. We present in this section the sub-procedure for genuine-token APTs as a representative, and leave the other two to Appendix~\ref{rm:APT_Construction}. Note that coin and genuine-token APTs are also called \textit{dust-value transfers} (DVTs). Moreover, a coin APT is always forward and active, and a backward genuine-token APT always has a zero value. Our two lists of genuine and fake tokens are described in Section~\ref{sec:data_collection}. Each fake token has a symbol that resembles that of a genuine token, referred to as its genuine counterpart.

    \begin{algorithm}
        \caption*{Genuine-Token APT Dataset Construction}
        \begin{algorithmic}
            \STATE\textbf{Step 1.} Identify the set $\tsg$ of \textit{suspicious} genuine-token transfers with values below 3 USD (at the time) in which \textit{both} senders and receivers are EOAs and not identical.
            \STATE\textbf{Step 2.} From the set $\tsg$ of suspicious transfers identified in Step 1, construct the set $\tagg$ of \textit{genuine-token} APTs and label them as follows. For each transfer $t=(s, r) \in \tsg$, first check if it is a \textit{forward} APT, by searching the transaction history of $r$. If there was a nonzero transfer of the same token $\ell=(r,b)$ that occurred before $t$ such that $b \neq s$ but $b$ and $s$ share at least three first and last digits, then include $t$ in $\tagg$ as a \textit{forward} APT with $s$ as the phishing address. If $t$ doesn't satisfy that condition, and $\v(t)=0$, then search in the transaction history of $s$ for a nonzero-value transaction $\ell = (s, b)$ that occurred before $t$ and $b \neq r$ but $b$ and $r$ share at least three first and three last characters. If such $\ell$ is found, include $t$ in $\tagg$ as a \textit{backward} APT with $r$ as the phishing address. In either case, we refer to $\ell$ as the \textit{legitimate transfer} associated with $t$ and $b$ its \textit{benign} address. Finally, if the phishing address is also the sender of the normal transaction of the same hash as $t$ then $t$ is an \textit{active} APT. Otherwise, it is \textit{passive}, and the sender of the normal transaction is called the \textit{coordinator} of the APT. 
        \end{algorithmic}
    \end{algorithm}
        
We find that among 3,855,694 coin APTs, 82.7\% are nonzero-value APTs.
Among 14,799,949 genuine-token APTs, 81.9\% are zero-value,
80.3\% are backward, and 83.5\% are passive.
Among 37,708,070 fake-token APTs, 99.7\% are backward, and almost
100\% are passive. See Appendix~\ref{app:APT_datasets} for more analysis.

\subsection{Identifying the Victims of APTs}
\label{subsec:victim_transfer}
Identifying victim transfers enables us to quantify the financial impact of APT scam and avoid mixing up scam-funders and victims.  

\begin{definition}[Victim Transfer]
    \label{def:victim_transfer}
    Let $\ta$ be a set of APTs.
    For each $t \in \ta$, denote by $p = \ph(t)$ and $a = \tg(t)$ the phishing address and the target address of $t$, respectively.  
    We refer to $P \triangleq \{p\colon p = \ph(t) \text{ for some } t \in \ta\}$ as the set of \textit{phishing addresses} and $A \triangleq \{a\colon a = \tg(t) \text{ for some } t \in \ta\}$ as the set of \textit{targeted addresses}. A transfer $f=(v,p)$ of a coin or a genuine token is called a \textit{victim transfer} (and $v$ is the  \textit{victim address}) if $f \notin \ta$ and either of the following conditions is satisfied, with (C1) taking precedence if both are satisfied.
    \begin{enumerate}[label=(C\arabic*), leftmargin=*, align=left]
        \item There was an APT $t \in \ta$ with $v=\tg(t)$ that occurred before $f$. This is a standard scenario where $v$ was targeted by an APT from $p$ and then mistakenly transferred funds to $p$ (see Fig.~\ref{fig:DVT_victim}b and Example~\ref{ex:68_0_DVT}). This is a `Type-1' victim transfer.

        \item Before $f$, there were an APT $t=(p,a) \in \ta$ and the associated legitimate transfer $\ell = (a,b)$ with $b\approx p$, and also a positive-valued non-APT transfer $\ell' = (v,b)\notin \ta$, which could occur at any point. This is a more subtle scenario when the victim used two addresses $a$ and $v$ to transfer funds to a common address $b$, and the address $a$ was the target of an APT, whereas $v$ was the address that actually sent fund out to $p$ (see Fig.~\ref{fig:DVT_victim}c and Example~\ref{ex:victim_versus_attacked}). This is called a `Type-2' victim transfer.
    \end{enumerate}
    We use $T_{V}$ to denote the set of victim transfers, and $V\triangleq \{v\colon \exists \ell=(v,p)\in T_{V}\}$ for the set of victim addresses. One can analyse $T_{V}$ to tally the total loss attributed to the APT scam.
\end{definition}

        We build the victim transfer dataset by finding all victim transfers going into each phishing address using the procedure \textsf{IsVictimTrans\-fer}, which implements Definition~\ref{def:victim_transfer}.

    \begin{algorithm}[htb!]
    \caption*{\textbf{IsVictimTransfer}$(p, f=(v,p), \tpa)$:}
        \begin{algorithmic}[1] 
        \STATE{\textbf{Input}: $p, f=(v,p), \tpa$ - the phishing address, a positive-valued non-APT incoming coin/genuine-token transfer, and the set of all APTs of $p$, respectively;}
        \FOR{$t \in \tpa$ that occurred before $f$}
            \IF{$\tg(t)=f$}
                \RETURN{TRUE;} \textcolor{gray}{\scriptsize {\#~type~1 victim transfer;}}
            \ENDIF
        \ENDFOR
        \FOR{$t \in \tpa$ that occurred before $f$}
            \STATE{$b \la \bn(t)$;} \textcolor{gray}{\scriptsize {\#~the address that looks similar to $p$ (see Section~\ref{subsec:APT_datasets});}}
            \IF{there exists a positive-valued non-APT transfer from $v$ to $b$}
                \RETURN{TRUE;} \textcolor{gray}{\scriptsize {\#~type~2 victim transfer;}}
            \ENDIF{}
        \ENDFOR
        \RETURN{FALSE;} \textcolor{gray}{\scriptsize {\#~not a victim transfer;}}
        \end{algorithmic}
    \end{algorithm}

    {\rev{We identified 4,209 victim transfers (4,180 type-1 and 29 type-2), in which 327 \ETH transfers (322 normal and 5 internal) and 1,505 genuine-token transfers have (historical) values of at least \$1,000 USD, corresponding to \$12.21 million in \ETH and \$153.6 million in tokens. Among these are the well-known cases of 20 million \USDT (frozen by Tether) in 2023 and 1,155 \WBTC in 2024 discussed in Example~\ref{ex:68_0_DVT}.}}

    \section{APT Masters and Distributors}
    \label{sec:master_distributor_datasets}

    We now shift our focus to APT master addresses, which execute or fund APTs, as well as their funders (called distributors). Masters play a central role in our study as the way they operate reveals significant insights into how the scammer organises their APT scams at scale (see Section~\ref{sec:signature_extraction}). We first define three typical master types 
    and then propose a systematic way to find them by pairing \textit{each} APT with a set of masters supporting it. Such fine-grained APT-masters pairings allow us to capture a specific type of scam signature called funding-residual signature (see Section~\ref{subsec:sig_funding_diff}), and provides an accurate way to connect `tainted' TC depositors and withdrawers with their associated masters (see Section~\ref{sec:TC_usage} and Appendix~\ref{app:TC_grouping}).

    \subsection{Three Master Types}
    \label{subsec:master_types}

    We define \textit{master} addresses (or masters for short) as the non-service EOA addresses that funded the phishing addresses (\textit{coin funders} or \textit{token funders}) or coordinated the APTs via contracts (\textit{coordinators}). 
    We first present typical ways an APT can be carried out and illustrate the different roles masters play in each case. Note that each phishing address can be involved in multiple types of APT and interact with different masters. The masters governing each of the 11 APT subtypes are shown in Fig.~\ref{fig:APT_subtypes_masters}. Various examples of master-phishing interactions can be found in Appendix~\ref{app:master_distributor_datasets}. 
    
    \noindent\textbf{Coin APT.} Each coin APT has at least one coin funder. 
    Typically, one coin funder supplied sufficient \ETH for the phishing to cover the APT amount \textit{and} the transaction fee.

    \begin{figure}[htb]
        \centering
        \includegraphics[width=\linewidth]{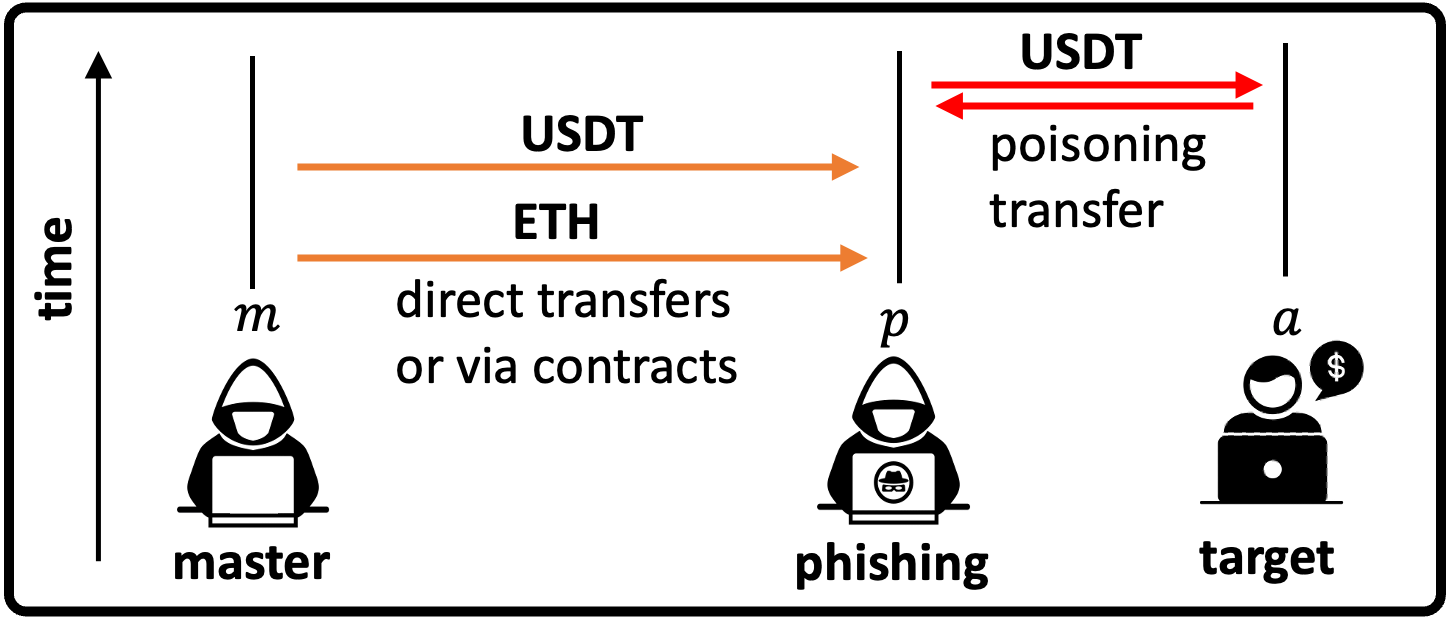}
        \caption{Funding for an \textit{active} genuine-token APT.}
        \label{fig:gentokenAPT_funding_active}
    \end{figure}

    \noindent\textbf{Genuine-Token APT.} For each genuine-token APT, the phishing address may need a coin funder, a token funder, or both, or none, depending on the specific APT subtype (see Fig.~\ref{fig:APT_subtypes_masters} for more details).  Fig.~\ref{fig:gentokenAPT_funding_active} shows the typical funding mechanism of an active, nonzero-value genuine-token APT, in which both coin and token funders are required (they can be identical). This is because in an active APT, the phishing address calls the transaction itself and hence needs coins to cover its fee. Furthermore, if it transfers a nonzero amount of genuine tokens then it also requires a token funder.   

    \begin{figure}[htb]
        \centering
        \includegraphics[width=\linewidth]{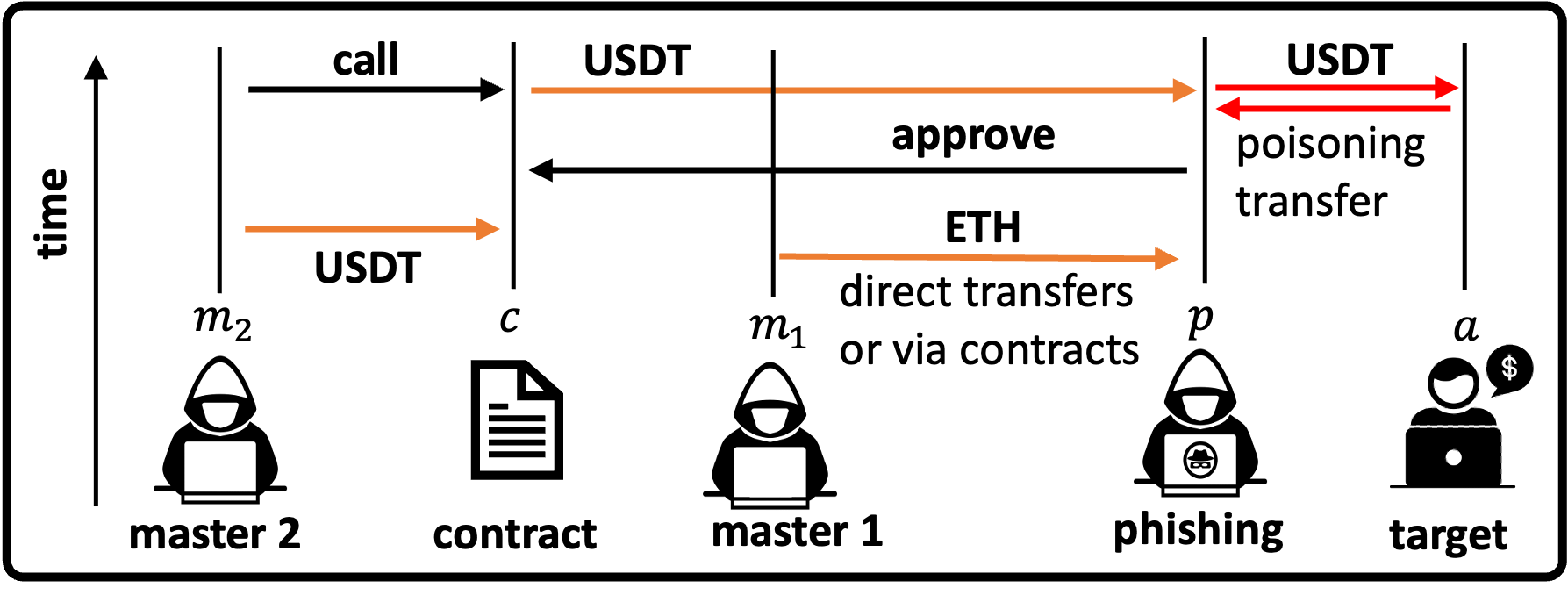}
        \caption{Funding for a \textit{passive} genuine-token APT.} 
        \label{fig:gentokenAPT_coordinator}
    \end{figure}

    Another common way for scammers to perform genuine-token APTs is to let a master $m_2$ (\textit{coordinator}) perform multiple APTs in the same transaction by calling a contract $c$ as depicted in Fig.~\ref{fig:gentokenAPT_coordinator}. For a nonzero value APT, another master $m_1$ (\textit{coin funder}) must first transfer some coin to the phishing address $p$ so that it can approve $c$ to spend the token on its behalf. Then, in a single contract call, the contract transferred some token to $p$ and then immediately transferred that from $p$ to the target. In this case, $m_2$ plays the roles of both \textit{token funder} and \textit{coordinator} for APT. Another possibility is to have a separate token funder $m_3 \neq m_2$. Note that in a forward zero-value genuine-token APT, a coin funder is not needed because a zero transfer does not requires approval from $p$. The same goes for a backward genuine-token APT.

    \begin{figure}[h]
        \centering
        \includegraphics[width=\linewidth]{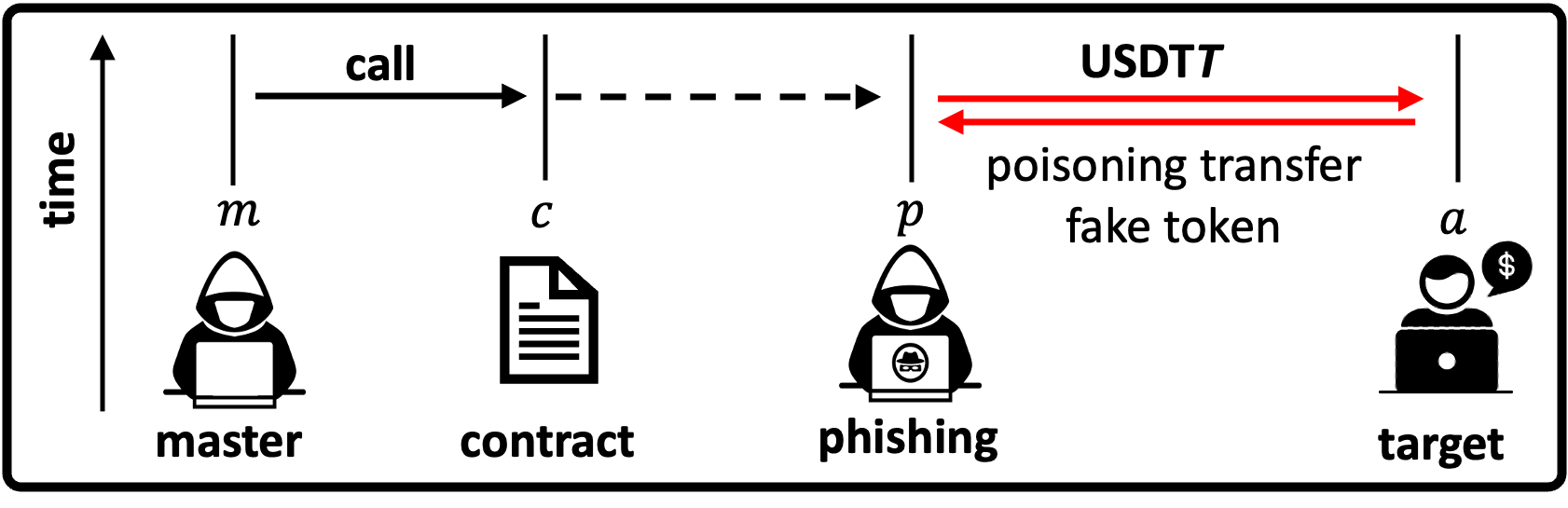}
        \caption{Typical operations for a passive fake-token APT.} 
        \label{fig:faketoken_APT_coordinator}
    \end{figure}
    
    \noindent\textbf{Fake-Token APT.} In the typical fake-token APT, a master (\textit{coordinator}) calls a contract that performs one or more transfers of fake tokens from phishing addresses to target addresses (see Fig.~\ref{fig:faketoken_APT_coordinator}). No funding to the phishing addresses is required.  

\subsection{Identifying the APT Masters}
\label{subsec:master_dataset}

    We developed a systematic approach to identify all master addresses and transactions associated to a given APT dataset (Section~\ref{sec:APT_datasets}). 
    For each phishing address $p$, we scan its list of APTs from oldest to newest and find all master addresses and transactions involved. Note that for each of the 11 APT subtypes, we know exactly the types of master we are looking for (see Fig.~\ref{fig:APT_subtypes_masters} and Section~\ref{subsec:master_types}). 
    
    Masters that are \textit{coordinators} can be easily found by retrieving the EOA addresses that call the transactions containing the APTs. For \textit{coin funders} and \textit{token funders}, we match each APT with one or more nearest incoming nonzero non-APT (coin/genuine token) transfers that occur before the APT and collectively cover its cost including transaction fee. The (non-service) EOA senders/transaction callers of these funding transfers are the master addresses. We allow partial funding, i.e. the same funding transfer can be used to fund multiple APTs with partial amounts. Our matching logic can handle complex cases including  multiple funding transfers that are used for multiple APTs (Example~\ref{ex:genAPT_multiple_coin_funders}), and victim transfers that are used to cover the cost of APTs (Examples~\ref{ex:funding_using_victim_transfers},  ~\ref{ex:funding_using_victim_transfers_ETH}). 

    We found 57,558,073 master records, corresponding to 5,307,343 unique transaction hashes, among which 2,780,856 transactions are for coin funding, 510,935 transactions are for genuine token funding, and 2,420,269 transactions are for APT coordination. Notable masters include \alinkLine{0x49c6246ab5209733d2ed59af17f9081d6f138370} who performed 12,639 transactions as \textit{coordinator}, supporting 189,645 phishing addresses, and \alinkLine{0xeFb3729Bc9C0ee64c718185d612C9538C0323306}, involved in 36,651 transactions as a coin/token funder, or coordinator, supporting 162,479 phishing addresses.
    
    There are 51,414 unique master addresses, 43,906 of which involved in one or two APT-funding/coordinating transactions. Many among them turn out to be phishing addresses. A random check shows that most belong to a long phishing chain (e.g. the one containing \alinkLine{0x6A968Fe887434Ba54dB54Ceca38A680A80f61224}), likely run by the same scammer given the chain topology and identical gas signatures. We henceforth focus on 7,508 masters that were involved in at least three APT-coordinating/funding transactions.

    We also define \textit{distributor}, which funded ETH to masters, and their funders (called \textit{distributor funders}). Examining the distributor dataset, we found a number of public services used to fund APT masters, led by FixedFloat (funded 291 masters), ChangeNOW (132 masters), and WhiteBIT (39 masters). Distributors and distributor funders also play important roles in our approach for grouping  `tainted' TC depositors/withdrawers as intermediate addresses between such TC users and the masters (Section~\ref{sec:TC_usage}). More details about distributors and their funders are in Appendix~\ref{app:master_distributor_datasets}.

   \section{APT Scam Signatures}
    \label{sec:signature_extraction}

    \rev{We formulate and study \textit{five} groups of scam signatures tailored to address poisoning transfers.} These signatures characterise the gas setting, the public service usage, the contract generation and usage, the relation between the APT values and benign values, and the relation between the APT values, transaction fees, and funding amounts. \rev{Such signatures capture different aspects of scam operations and allow us to gain deeper understanding of how scammers run their campaign, most likely by using a scam toolkit or automated software. We note that Tsuchiya {\et}~\cite{tsuchiya_USENIX_2025} briefly discussed the idea of scam signatures using terms like `attack strategies', `behaviours', `contract reuse',  without going into concrete details, except for the observation that one of the clusters used contracts with identical bytecodes. This section presents a detailed study of such behavioural patterns observed from the APT datasets.} 

   We evaluate scam signatures using two complementary metrics. Address-level consistency measures how consistently an address reuses the same signature values, while signature-level cohesion measures the extent to which same-signature address groups are corroborated by scam-operation evidence.

    \subsection{Consistency Scores for Scam Signatures}
    \label{subsec:normalised_consistency_scores}
    We define the Inverse-Simpson/Shannon \textit{consistency scores} to quantify how consistently an address reuses observed elements (e.g. contracts) relative to the number of scam-related transactions. The closer the score is to one, the more consistent the behaviour, i.e. using fewer values with larger frequencies. A more consistent signature reflects the behaviour of an address more faithfully. Details are provided in Appendix~\ref{app:consistency_scores}.

    \subsection{Cohesion Score for Scam Signatures}
    \label{subsec:cohesion_score}

    The cohesion score is developed based on the \textit{guilt-by-association (GBA) graph}, which `connects' two masters if they support the same scammer address, or fund each other, or are funded by the same non-service distributor. As a common assumption in the literature (see, e.g. \cite{GuanLi_CCS24, tsuchiya_USENIX_2025, Huynh_etal_WWW2025}), addresses that lie in the same weakly connected component of such graphs likely belong to the same scammer group. The concrete definition of the GBA graph can be found in Appendix~\ref{app:cohesion_score}.

        We define cohesion at both the group and signature levels. For a group of masters sharing the same signature value, \textit{group cohesion} is defined as the fraction of master pairs that lie in the same WCC of the induced GBA graph. \textit{Signature cohesion} is defined similarly but over all such groups (of size at least two). The highest cohesion score is one, which means that \textit{every} pair of masters that have the same signature value can be \textit{connected by a path} in the corresponding GBA graph. The lowest cohesion score is zero, which means that none of the pairs of masters with the same signature value are connected in the corresponding GBA graph. The formal definition of signature cohesion can be found in Appendix~\ref{app:cohesion_score}.

    \subsection{Gas Signatures}
    \label{subsec:sig_gas}

    The sender of an Ethereum transaction can specify a few gas-related parameters such as the \textit{gasLimit}, which is the maximum gas provided by the sender, the \textit{maxFeePerGas} (only for transactions of type-2, type-3), which is the maximum price (in Gwei) they are willing to pay for each gas unit, and the \textit{maxPriorityFeePerGas} (only for transactions of type-2, type-3), which is the tip (in Gwei) they are willing to pay to the block builder per gas unit. These parameters, which are set by the scammer, can be used as scam signatures.

    \begin{definition}[Gas Signatures]
    \label{def:sig_gas}
        We define three \textit{gas signatures} $\sgl$, $\smfpg$, and $\smppg$ for each address $a$, corresponding to gasLimit, maxFeePerGas, and maxFeePerPriority as follows. Each signature is a multiset $\{(v_1,f_1),\ldots,(v_k,f_k)\}$, where the value $v_i$ appears $f_i$ times in the scam-related transactions of $a$.
    \end{definition}

    \begin{example}\label{ex:gas_sig_1}
        The master $m_1 = $ 
        \alinkLine{0xc16acf1380a144f673411226c9dd913cf1eec182} performed more than 8,000 transactions, each of which carries many APTs and always has maxFeePerGas 80 Gwei and maxPriorityFeePerGas 3 Gwei, but with gasLimit 8,500,000 in early transactions and 18,500,000 in later transactions. More precisely, $\sgl(m_1) = \{(18500000,7071), (8500000,1354)\}$, $\smfpg(m_1) = \{(80,8425)\}$, and $\smppg(m_1) = \{(3,8425)\}$.
        The master $m_2 = $ 
        \alinkLine{0xefb3729bc9c0ee64c718185d612c9538c0323306} has $\sgl(m_2) = \{(6000000, 31833), \allowbreak(500000, 3865)$, $(2000000, 403), (3000000, 601)\}$, $\smfpg(m_2)=\{(80,\allowbreak 36702)\}$, and $\smppg(m_2) = \{(3, 36702)\}$. 
     \end{example}

    \textbf{Gas Signatures Consistency}. We collect gas signatures restricted to APT-carrying transactions for all 7,508 master addresses that have at least three APT-related transactions. The consistency scores (Section~\ref{subsec:normalised_consistency_scores}) of these signatures are presented in Table~\ref{tab:gas_sig_ncs}. We found that $\smppg$ is the most consistent gas signature, with 65.9\% of master addresses using exactly one value across all transactions, and average consistency scores 0.81 (IS) and 0.8 (SH). Although $\sgl$ appears reasonably consistent, we note that among the 4,493 masters that used only one value for gasLimit, 2,085 used the default value of 21,000, which provides no useful insight. 

      \begin{table}[htb]
    \caption{Percentages of masters that have \textit{gas signatures} consistency scores of 1, at least 0.9, at least 0.8, and their mean. Each entry uses the format IS|Shannon.}
        \label{tab:gas_sig_ncs}
        \centering
        \small
        \begin{tabular}{@{}lcccc@{}}
             \toprule
             Sig/Score     & $=1$ & $\geq 0.9$ & $\geq 0.8$ & mean\\ 
            \midrule       
             GL           & 59.8|59.8 & 65.3|64 & 68.6|68 & 0.78|0.77 \\
             \midrule
             MFPG       & 36.6|36.6 & 38.3|37.8 & 39.7|38.9 & 0.42|0.42 \\
             \midrule
             \textbf{MPFPG}  & \textbf{65.4|65.4} & \textbf{77.6|72.5} & \textbf{79.2|77.5} & \textbf{0.81|0.8} \\  
            \bottomrule          
        \end{tabular}
    \end{table}

    \textbf{Gas Signatures Cohesion.} The gas signatures GL, MFPG, and MPFPG have cohesion scores \rev{0.58, 0.88, and 0.69}, respectively. This means, for example, that 88\% of the master pairs that have identical MFPG signature values are `connected' (i.e. having a path connecting them) in the corresponding scam-operation graph. These scores suggest that the gas signatures are reasonably effective in capturing master addresses that likely belong to the same scammer.

    \subsection{Contract Signatures}
    \label{subsec:sig_contract_calls}

    The \textit{contract signatures} capture the observation that each scammer tends to generate or use contracts with identical (or similar) codes for funding APTs, coordinating APTs, moving funds among different addresses, or for fake tokens. 
    In this section, we formulate the so-called \textit{contract-generation} and \textit{private-contract-call} signatures, and analyse these signatures over 7,508 master addresses.
    To avoid the expensive pairwise comparison of contract codes, we replace a contract address by the first 40 hex characters of the Keccak256 hash of its bytecode. Thus, contracts with identical bytecode will be referred to by a common (modified) address, eliminating the need for pairwise comparisons. For example, the APT contracts 
    \alinkLine{0x7C439FDeB9F835A34221B09599Fd85e83B1Cdf3b} and 
    \alinkLine{0x11972d7E8aB2274F342b6B340C91CcD50b0459eB} have identical bytecode, and hence both will be referred to using a new address 
    \alinkLine{0x40a8dc0fb84193a1d20e95e76ae98e90a81075d5} obtained from the hash of their bytecode. 

    \begin{definition}\label{def:modified_contract}
        A \textit{modified address} $c'=h(c)$ of a contract $c$ is obtained by concatenating \path{0x} and the first 40 hex characters of the Keccak256 hash of its runtime bytecode.
    \end{definition}

    \begin{definition}[Contract-Generation Signature] \label{def:sig_contract_generations}
        The \textit{contract-generation signature} $\scg(m)$ of an address $m$ is the set of modified addresses of the contracts created by $m$.
    \end{definition}
    
    \begin{definition}[Private-Contract-Call Signature] \label{def:sig_contract_calls}
        Given an address set $M$ and an APT/fake-token contract set $F$, the \textit{private-contract-call signature} $\scc(m)$ of $m \in M$ is the multiset of modified addresses of the contracts called by $m$ (with frequencies) that either are APT/fake-token contracts from $F$ or created by some $m'\in M$.

    \end{definition}

        \begin{example} \label{ex:master_contract_calls}
        The master \textbf{0xc16a} (see Example~\ref{ex:gas_sig_1}) created and called the APT contract 
        \alinkLine{0xb9f6a074d8093cb7c48c53d42776ca7dca6fda52} 8,425 times. Thus, its CG and CC signatures are $\{\text{0x40a8...75d5}\}$ and $\{(\text{0x40a8...75d5}, 8425)\}$, respectively, where \text{0x40a8...75d5} is the modified address of the contract. 
        The master 
        \alinkLine{0xA57188bD032cf7fA7319608f0FD9F155b387205D} generated and used six contracts, among which 
        \alinkLine{0x4b018b99d990825079c9794c382dfaa8bbd395a6} is an APT contract and was used 3,164 times, while the other five were used once (for setting the whitelist) and have identical bytecodes. By Definitions~\ref{def:sig_contract_generations} and~\ref{def:sig_contract_calls}, its CG and CC signatures are $\{\text{0xc527...c4b0}, \text{0x5f31...c6ab}\}$ and $\{(\text{0xc527...c4b0},3164),$ $(\text{0x5f31...c6ab},5)\}$,  where \text{0xc527...c4b0}/\text{0x5f31...c6ab} are modified addresses of the APT/whitelist-setting contracts. 
    \end{example}

    \textbf{Generation of Contract Signatures.} We generated the contract signatures for the list \rev{$M_{\geq 3}$} of all 7,508 masters that were involved in at least three funding/APT transactions (see Section~\ref{subsec:master_dataset}) as follows. First, we collected the transaction hashes of successful out-going normal transactions from all $m\in M_{\geq 3}$ that have an empty `\texttt{to}' field (i.e. contract-creation transactions). 
    We then called an API from Etherscan to collect the list $C(M)$ of \rev{13,750} corresponding contract addresses generated by such transactions (by \rev{3,472} masters). We also identified \rev{1,133} APT/fake-token contracts that were \textit{not} created by these masters (i.e. $F\setminus C(M)$). Another API was called to retrieve the runtime bytecodes of such contracts, which were then hashed using Keccak256 to produced \rev{14,883} modified addresses. 
    The CG signature of $m\in M$ is the set of modified addresses $h(c)$ for $c$ created by $m$, and its CC signature is the multiset of $h(c)$ for $c \in C(M)\cup F$ called by $m$.

    \textbf{Contract Signature Consistency.} 
    We identified \rev{4,112} masters that called \rev{6,176} private contracts $c\in C(M)\cup F$. Their CC signatures are highly consistent, with 76.5\% calling exactly one (modified) contract, and approximately 90\% having their IS|Shannon scores at least 0.9. Both scores have mean 0.95. Consistency scores do not apply to CG signature as each master creates each contract once.

    \textbf{Contract Signatures Cohesion.} CC and CG signatures have remarkably high cohesion scores, 0.959 and 0.998, respectively. This means that more than 95\% of pairs of masters that share the same contract signatures turn out to be connected by a path in their corresponding scam-operation graph. This signifies that contract signatures are highly effective for clustering master addresses controlled by the same scammer.

    \textbf{Signature-Based Address Grouping}. Based on the CG signature, we found that 3,415 master addresses that created some contracts contain 38 groups of sizes between 2 and 31, and four groups of sizes 1699 (CG: \{0x9029\ldots 649b\}), 467 (CG: \{0x5e66\ldots c3c3\}), 254 (CG: \{0x5f31\ldots c6ab, 0xc527\ldots c4b0\}), and 247 (CG: \{0x5e66\ldots c3c3\}), referred to as Groups CG1, CG2, CG3, CG4, respectively. These groups funded and coordinated large numbers of phishing addresses and APT transactions. For instance, CG1-CG4 supported 485948, 125247, 1090933, 1476714 unique phishing addresses, and 352890, 37383, 539626, and 135398 unique APT-carrying transactions, respectively. Each remaining 534 masters has a unique CG signature.   
    
    \textbf{Cross-Signature Analysis.} The above groups also demonstrate remarkably consistent gas signatures. In Group CG1, all masters used a single value of 3 for maxPriorityFeePerGas, 1643 out of 1699 masters used a single value of 200 for maxFeePerGas and 18,500,000 for gasLimit.  In Group CG2, all 467 masters used identical gas signatures, with corresponding values 3, 200, and 18,500,000. In Group CG3, 251 masters used 1.5 as the only maxPriorityFeePerGas, and the remaining three used both 0.1 and 1.5. Group CG4 is less consistent, with 149 masters using 1 and 73 masters using 1.5 as their single maxPriorityFeePerGas, and a few others use both 1 and 1.5, or both 0.1 and 1. In addition, CG1, CG2, CG3 have \textit{perfect} group cohesion scores of 1, whereas CG4 has a high cohesion score of 0.91, providing further evidence that each master group belongs to the same scammer. 

    Interestingly, contract and gas signatures enable us to formalize and substantiate the qualitative observation by Tsuchiya~{\et}~\cite{tsuchiya_USENIX_2025} that copying bots connect ``two seemingly distinct large groups (based on their strategies and behaviours)'', which might have led to a possibly incorrect address grouping in~\cite{GuanLi_CCS24} (see Appendix~\ref{app:copying_bot} for further details).

    \subsection{APT Funding Residual Signatures}
    \label{subsec:sig_funding_diff}

    In a coin APT, the master must first fund $x$ coins to the phishing address, which then performs an APT with value $y$ and transaction fee $z$ coins. 
    The multiset of differences $x-(y+z)$ forms the so-called funding-residual (FR) signature of the master address (Definition~\ref{def:sig_funding_diff}). This also extends to genuine-token APTs with a slight modification (remove the fee $z$ in the formula). To accurately collect the FR signatures, we must first pair the funding transfers and the APT transfers correctly, which is a nontrivial task (see Section~\ref{subsec:master_types}) that none of the previous work on APTs has investigated.

    \begin{definition}[Funding-Residual Signature]\label{def:sig_funding_diff}
        Suppose a master $m$ funded $x_i$ \ETH to its $i$-th coin APT, which has a value of $y_i$ \ETH and costs a transaction fee of $z_i$ \ETH, $i=1,\ldots,k$. Then, the funding-residual (FR) signature of $m$, denoted $\sfr(m)$, is the multiset of the residuals $x_i-(y_i+z_i)$, $i=1,\ldots,k$. 
    \end{definition}

    To use a multiset signature for address grouping, we first flatten them into sets with a threshold $\theta \in (0,1]$. Setting $\theta < 1$ eliminates outlier values due to unexpected changes in the fee market,  occasional incorrect pairings of APT-carrying and funding transactions, or numerical errors in computation. For example, for $\theta = 0.8$, both multisets $\{(1, 8), (2, 1), (3, 1)\}$ and $\{(1,10)\}$ are flattened to $\{1\}$.
    
    \begin{definition}[$\theta$-Flattened Multiset]\label{def:sign_funding_diff_flattened}
        For $\theta \in (0,1]$, the $\theta$-flattened set of a multiset $\left\{(v_i,f_i)_{i=1}^k\right\}$ (with $f_1\geq f_2 \geq \cdots \geq f_k$) is $\{v_i\}_{i=1}^{k'}$, where $k'$ is the smallest integer satisfying $\sum_{i=1}^{k'} f_i \geq \theta \sum_{i=1}^k f_i$. 
    \end{definition}

    \begin{example} \label{ex:sig_coin_diff_0}     
        The master 
        \alinkLine{0x111701b067Ab44B8c358514775ba5322b3628BCc} funded 
        0.000528850756707 \ETH to the phishing address 
        \alinkLine{0xB50B26e9681c6433fc82d106BA7D3969CE185564}, which then performed an APT of 0 \ETH and a transaction fee exactly the same as the funding amount: 0.000528850756707 \ETH. 
        The funding residual for this pair of APT and funding transactions is 0. In fact, this master has the FR signature $\{(0,4000), (v_1,1), (v_2,1),\ldots,(v_{118},1)\}$. According to Definition~\ref{def:sign_funding_diff_flattened}, its $0.8$-flattened FR signature is $\{0\}$. This FR value of 0 appears in 97\% APT-funding transactions.
        
    \end{example}

    Note that the FR signature applies to token APTs as well. For example, the master 
    \alinkLine{0x3826705213c38a060769b56f6880fd1f1ef999e9} always funded 0.00001 more \USDT/\USDC than the APT value. For instance, it funded 
    the phishing address 
    \alinkLine{0x4de97Dc7357823Cd601EcDd8f901DffCa9909416} 2.75002 \USDC, which in turn sent an APT of $2.75001 = 2.75002-0.00001$ \USDC. However, we do not investigate the FR signature for genuine tokens since from our observation most masters funded the exact token amounts needed for the APTs, making the residual 0 and the signature less useful. 

    \textbf{Funding-Residual Signature Consistency.} We measured the Inverse-Simpson and Shannon consistency scores for the FR signatures of \rev{3,535 master addresses that funded 3,806,347 coin APTs, and found that 34.3\% have perfect consistency scores of 1, 
    91.3\%|81.8\% have scores at least 0.8, and with mean scores 0.9|0.87 for the IS|SH scores. Notable examples of masters with FR signature consistency at least 0.99 include 
    \alinkLine{0xdbdcef4368d35267753a9958f965b24d60b47613} (funded 31,741 coin APTs, 99\% with FR value $0$ \ETH), 
    \alinkLine{0x1300080e2a951be01a5a87c6897bf59e65118013} (funded 16,827 coin APTs, 99\% with FR value $9.99\times 10^{-8}$ \ETH), 
    \alinkLine{0x760cd3aa47f434887e75ff40fef466faf037f4df} (funded 12,840 APTs, 99\% with FR value $10^{-14}$ \ETH), and 
    \alinkLine{0x700c030143f08862eb64263679230b8bec00f699} (funded 3873 APTs, 99\% with FR value $7.9\times 10^{-14}$ \ETH).}

    \textbf{Funding-Residual-Signature Address Grouping.} We first flatten the multiset that represents the FR signature of each master to a set of funding residual values and then form groups of master addresses that have similar or identical signatures using a dictionary with signatures as keys and the correlated masters as values. We use a threshold percentage $\theta$ when flattening the multisets to eliminate infrequent values. Setting $\theta=80\%$, the set of \rev{3,535} master addresses who served as coin funders of coin APTs can be partitioned into 250 groups. The four largest groups are: Group FR1 with \rev{913} addresses (flattened FR signature $\{0\}$), Group FR2 with \rev{344} addresses ($\{9.99\times 10^{-8}\}$), Group FR3 with \rev{216} addresses ($\{10^{-18}\}$), and Group FR4 with 90 addresses ($\{2.1\times 10^{-9}\}$). These groups funded \rev{407462, 400907, 72250,} and 28641 unique phishing addresses, respectively. 
    Note that an address having $0.8$-flattened FR signature $\{r\}$ means that the funding residual is exactly $r$ in at least 80\% of its funding transactions. 

    \textbf{Funding-Residual Signature Cohesion.} (Flattened) FR signature has a low cohesion score of \rev{0.52}. The main reason is that the largest group of 913 masters FR1, which has signature $\{0\}$, has cohesion score of only 0.42. Our hypothesis is that the trivial value of $0$ can be used by multiple scammers, which means that FR1 may consist of addresses from different scammers, which are not connected in the scam-operation graph. Further evidence supporting this hypothesis is discussed in the analysis of the Value-Pattern signature in the next section. On the other hand, removing FR1 from the calculation pushes the cohesion score to \rev{0.97}, which suggests that FR values other than 0 are more reliable for master address grouping.

    \subsection{APT-Value-Pattern Signatures}
    \label{subsec:sig_APT_benign}

    We observe from the coin and genuine-token APT datasets that APTs funded by the same master address frequently exhibit either constant transfer values or values that systematically resemble corresponding benign transfers. For example, a benign value of 12345 may be met with an APT value of 0.012345, suggesting a deliberate transformation that preserves the digit sequence while altering scale. The intent is to mirror visual similarity between benign and malicious transactions when the decimal point is ignored, thereby enhancing the likelihood of deceiving the victim. To capture this class of signatures, a generic approach is to collect all (APT value, benign value) pairs associated with APTs funded by a given master address and derive a compact benign-to-APT mapping. 
    
    We now formalise and evaluate the usage of two mappings: APT values are constant, or visually match the benign values while staying in a value range , e.g. $[10^{-5},10^{-4})$. Such strategies potentially allow better control of the APT costs.

    \begin{definition}[Constant-VP Signature]\label{def:constant_VP}
        The Constant-APT-Value-Pattern (c-VP) signature of a master address $m$ is the multiset of strings `CONST\_$a$' with their frequencies, where each string `CONST\_$a$' is obtained from an APT with value $a$ funded by $m$.
    \end{definition}

    \begin{definition}[Mimic-VP Signature]\label{def:mimic_VP}
         Given an integer $d\geq 1$, the $d$-Mimic-APT-Value-Pattern (m-VP) signature of a master address $m$ is the multiset of strings `MIMIC\_$d$\_$E$' and `OTHER' obtained from APTs funded by $m$ as follows. For each pair $(a,b)$ representing (APT-value, benign-value), if the $d$ left-most digits of $a$ and $b$ after the scaling transformation (removing their decimal point and the 0's on the left and right, and adding 0's to the right if the number of digits falls below $d$) do not match, then produce the string `OTHER'. Otherwise, set $E = \lfloor \log_{10} (a)\rfloor$. 
    \end{definition}

    We note that the collected pairs (APT-value, benign-value) may have \textit{noise} since an APT can sometimes be paired with a benign transfer that is algorithmically correct but was not the one intended by the scammer. We use a threshold $\theta$ to eliminate infrequent noise when aggregating the two signatures.

    \begin{definition}[VP Signature]\label{def:VP_sig}
        For $\theta \in (0,1]$, the $\theta$-Aggregated-APT-Value-Pattern (VP) signature of a master address is a single string of one of the forms `CONST\_$a$', `MIMIC\_$d$\_$E$', or `OTHER', obtained as follows. Let `CONST\_$a$' and `MIMIC\_$d$\_$E$' be the two strings with largest frequencies in the c-VP and the m-VP signatures of $m$, and $r_{\text{C}}$ and $r_{\text{M}}$ be the ratios of their frequencies versus the total number of APTs funded by $m$, respectively. Then the VP signature of $m$ is `OTHER' if $\theta > \max\{r_{\text{C}}, r_{\text{M}}\}$. Otherwise, it is `CONST\_$a$' if $r_{\text{C}} \geq r_{\text{M}}$, or `MIMIC\_$d$\_$E$' if $r_{\text{C}} < r_{\text{M}}$.
    \end{definition}

    \begin{example}
    \label{ex:apt_map_const}
        The master 
        \alinkLine{0x8fe6b2257DD8aF5f4B83a497C092F641Dd22F6b1} funded 5577 coin APTs, 4541 (81\%) among which used a constant value of 0.00001 \ETH regardless of the benign values (e.g. 
        \txlinkLine{0x18cb1d982339b6b3cb042dc64afdc7e221d6ebb4a097ac0be8d30cb4e65f8dcd}). It also used `MIMIC\_6\_$-5$' for 1775 (32\%) APTs. For example, the APT \txlinkLine{0xf2105ce09c1966728e68be3801eff71d809ac9455f1b1f47c5200f1ed8deb015} has value 0.000016 \ETH to mimic the benign value 16 \ETH of the transfer \txlinkLine{0x5315c88faf015fbd5e997d8b2e977ac9155b3f36239c980ef4de7b08b2a9c43b}. According to Definition~\ref{def:VP_sig}, this master has VP signature `CONST\_0.00001' with $\theta=0.8$. 
    \end{example}

    \textbf{VP Signature Consistency.} We examined \rev{3,535} masters that funded coin APTs and found \rev{17} groups (summing to \rev{2,682} addresses) that have VP signatures (other than `OTHER') with the threshold $\theta = 0.8$. This means that \rev{76\% of the 3,535} master addresses of interest exhibited the same VP signatures in at least 80\% of their APTs. The top groups include: \rev{VP1 with 1,238 addresses and VP signature `CONST\_0' (funded 522,731 APTs)}, VP2 with 605 addresses and VP signature `MIMIC\_6\_$-6$' (funded \rev{115,046} coin APTs), VP3 with \rev{443} addresses and VP signature `CONST\_$10^{-10}$' (funded \rev{637,309} coin APTs), VP4 with \rev{308} addresses and VP signature `CONST\_$10^{-6}$' (funded \rev{1,544,033} coin APTs). Other notable VP signatures include `CONST\_$a$' with $a \in \{10^{-18}, 10^{-4}, 10^{-5}, 10^{-8}\}$ (\rev{65} addresses in total), and `MIMIC\_6\_$-4$' (17 addresses).

    \textbf{VP Signature Cohesion.} \rev{VP signature
    has a low cohesion score of 0.24. The main reason is that the largest group VP1 (1,238 masters) has cohesion score of only 0.003. Our guess is that $0$-value APTs are commonly used even among independent scammers due to its low cost. Removing VP1, however, increases the cohesion score to 0.79, which suggests that other VP values could still be useful for identifying master addresses likely from the same scammer group.}
 
    \textbf{Cross-Checking with Funding-Residual Signatures.} Overlaps among some large groups of the two signatures are noticeable. 
    For example, Group FR1 contains 595 out of 605 addresses from VP2, but also 93 from VP3 and 76 from VP1, suggesting that FR1 may not belong to a single scammer group. Group FR2, however, lies entirely inside VP3, and also has a perfect cohesion score of 1, strongly suggesting that its addresses could be run by the same scammer group (example master address: \alinkLine{0x309eecd11dd36c69ffbdfcc54a20ffb639a6600c}).

    \subsection{Public-Service Signature}
    \label{subsec:sig_public}

    The \textit{public-service} signature lists the addresses 
    from public services used by the scammer. It is plausible that the same scammer may have a list of preferred services to interact with. 

    \begin{definition}[Public-Service Signature]\label{def:sig_public}
        The \textit{public-service} signature $\spub(a)$ of an address $a$ is the multiset of public service (EOA and contract) addresses with which $a$ interacted. 
    \end{definition}

    \begin{example}
        The distributor 
        \alinkLine{0x5a75cd71cE5eaC048a4b1479abc98A804c828c87} (see Fig.~\ref{fig:example_DVT_network_TC}) used only 1inch's Aggregation Router V5 
        \alinkLine{0x1111111254EEB25477B68fb85Ed929f73A960582} to fund 25 master addresses and did not interact with other services. Therefore, its public-service signature is $\{(0x1111...0582, 25)\}$. One of its successful phishing addresses \textbf{0x4f51} (also mentioned in Fig.~\ref{fig:example_DVT_network_TC}) interacted with just the Tornado Cash Router\alinkLine{0xd90e2f925DA726b50C4Ed8D0Fb90Ad053324F31b}. In fact, these two public services were repeatedly used in this scam group, forming a distinctive mechanism to fund, reinvest or launder scam proceeds.
    \end{example}

    \section{Usage of Tornado Cash in APT Operations}
    \label{sec:TC_usage}

    \rev{In this section, we provide an estimate of the total APT scam proceeds laundered via public services. We also investigate the Tornado Cash user addresses that are involved in APT scam operations with various roles such as phishing, master, distributor, laundering funder, and laundering addresses. Such addresses contributed to approximately {\red{\$38M}} deposited into and withdrawn from Tornado Cash (11/2022-5/2025).}

    \subsection{Laundering via Public Services}
    \label{subsec:flow_to_TC}

    To estimate APT scam proceeds laundered into public services, we propose a graph traversal algorithm called Dynamic Laundering Flow (DLF), which builds the collection of transfers that likely carry the scam proceeds within a bounded number of hops from the victim addresses. At its core, the algorithm has two distinctive features compared to standard BFS-based tracing algorithms (see e.g. \cite{Yan_etal_Cybersecurity_2023, Wu_etal_TIFS_2024}). First, it follows the \textit{time-value constraint} that the value of each outgoing laundering transfer should not exceed the total value of all previous incoming laundering transfers minus the total value of all previous outgoing laundering transfers. Second, it will \textit{revisit} an address and \textit{reprocess} its outgoing laundering transfers if it identifies new incoming laundering transfers. The most closely related money-laundering tracing algorithm to ours is MFTracer by Huo {\et}~\cite{HuoHuZhouYuWuWang_MFTracer_SigMetrics_2025}. However, as MFTracer converts the cryptocurrency flow into USD to simplify the tracing, it suffers from temporal valuation errors due to fluctuations in token prices (see~\cite[Sec.~5.3]{HuoHuZhouYuWuWang_MFTracer_SigMetrics_2025}). By contrast, our algorithm tracks individual tokens, and also handles token swaps via DEX. Further details can be found in Appendix~\ref{app:laundering}.
    
    Applying DLF to victim transfers of at least {\rev{\$1,000}} and tracing up to {\rev{15 hops}} from the victim addresses, we identified {\rev{17,566.99}} \ETH (about {\rev{\$38.09 million}} at the time of transfer) in APT scam proceeds entering TC. This represents {\rev{22.97\%}} of the total observed victim losses—approximately {\rev{\$165.82 million}} across {\rev{1,832}} victim transfers between November 2022 and May 2025. The TC 100 \ETH pool received approximately {\rev{91.62\%}} of the observed TC inflow. APT scam proceeds also flowed into other services (see Table~\ref{tab:laundering_public_services} and Appendix \ref{sec:anatomy_of_the_laundering_networks}). 

\begin{table}[h]
    \centering
    \small
    \caption{Estimates of APT scam proceeds flowing into public services after running DLF for \rev{15 hops}.}
    \label{tab:laundering_public_services}
    \begin{tabular}{@{}lrr@{}}
    \toprule
    Service & Amount (\$M) & Share (\%) \\
    \midrule
    Tornado Cash & 38.09 & 67.735 \\
    Avalanche & 4.83 & 8.595 \\
    SWFT Swap & 3.39 & 6.026 \\
    FixedFloat & 3.38 & 6.016 \\
    eXch & 2.33 & 4.139 \\
    Others & 4.21 & 7.488 \\
    \midrule
    Total & 56.23 & 100.00 \\
    \bottomrule
    \end{tabular}
\end{table}

\textbf{Laundering Funders.} We identified 944 \textit{laundering funders}, who funded 1,525 successful phishing addresses (that received a victim transfer of at least \$1,000) 
For example, the laundering funder \alinkLine{0xB1a2b6d0D19291C5354fe034B33f1e98D7f4444A} transferred 0.0277 \ETH to the phishing address \alinkLine{0x67694AF0ee15792a89573c72cDd21e4560d375f7} so that it can transfer 300,000 \USDT received from a victim.
\subsection{Tornado Cash User Addresses Related to Address Poisoning}
\label{subsec:TC_grouping}

    \textbf{Identifying Tainted TC User Addresses.} `Tainted' depositors are found by the laundering tracing algorithm DLF, whereas tainted withdrawers are those that also play one of the defined roles in APT operation, including phishing, masters (see Section~\ref{subsec:master_dataset}), distributors and distributor funders (see Section~\ref{subsec:master_dataset} and Appendix~\ref{app:distributor_dataset}), and laundering funders (see Section~\ref{subsec:flow_to_TC}).

   We identified 85 tainted depositors, which include 7 phishing, 4 masters, 9 distributors, 14 distributor funders, and 4 laundering funders (note that one address may be assigned several APT roles). We also caught 138 tainted withdrawers that also play the role of phishing (3), master (16), or distributor (32), distributor funder (87), and laundering funder (27).
   Addresses from this group have deposited 20030.3 ETH across 476 transactions and withdrawn 1967.3 ETH across 233 transactions.

\begin{figure}
       \centering
       \includegraphics[width=1\linewidth]{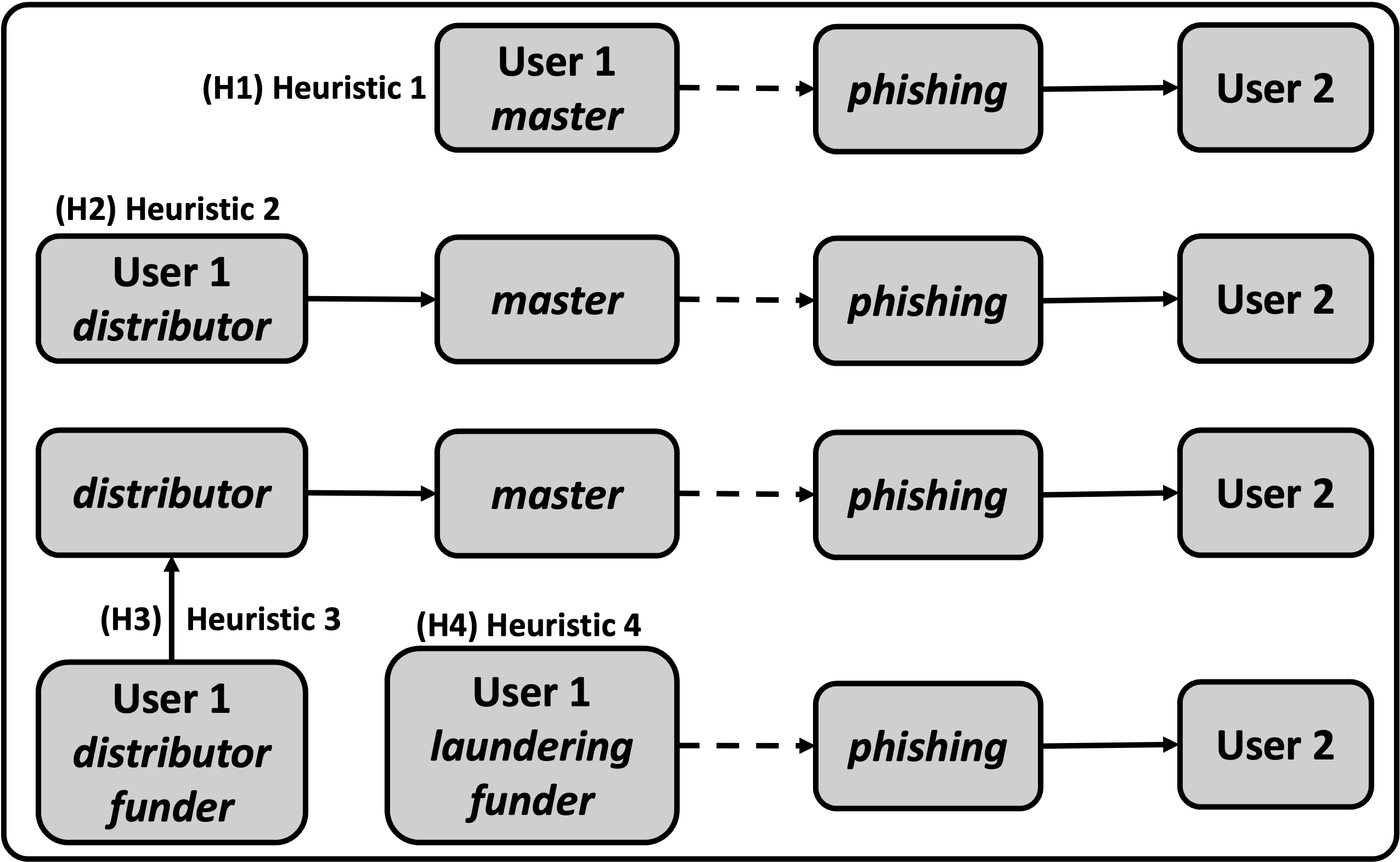}
       \caption{APT-based heuristics for grouping addresses. Solid arrows refer to (nonzero, non-APT, non-self-transfer) asset transfers, whereas dashed arrows refer to coin/token funding or coordination of APT.}
       \label{fig:TC_grouping_heuristics}
   \end{figure}

   \textbf{APT-based Clustering Heuristics.} We introduce four clustering heuristics based on the insight of APT operations (see Fig.~\ref{fig:TC_grouping_heuristics}).
   \begin{enumerate}
       \item (H1) User 1 is a master that coordinated/funded an APT involving a phishing that funded User 2.
       \item (H2) User 1 is a distributor that funded a master that coordinated/funded an APT involving a phishing that funded User 2.
       \item (H3) User 1 is a distributor funder that funded a distributor that funded a master that coordinated/funded an APT involving a phishing that funded User 2.
       \item (H4) User 1 is a laundering funder that funded a successful phishing address that funded User 2.
   \end{enumerate}
   We also use two common heuristic that prove to be most useful in previous works such as~\cite{WangChaliasosQinZhouGaoBerrangLivshitsGervais_WWW_23}: (H5) - User 1 and User 2 have a direct asset transfer, and (H6) - Two depositors have a common funder. 

   \textbf{Findings.} We have found 16 connections using (H1), 380 connections using (H2), 49 connections using (H3), 91 connections using (H5), and 53 connections using (H6). Note that these do not necessarily correspond to unique pairwise connections.  Using (H5) and (H6) alone, we found 182 groups (connected components) among all tainted depositors and withdrawers, with sizes 17, 6, 5, 4, 4, etc. 
   Using our heuristics alone, we obtained 174 groups of sizes 16, 11, 9, 4, 4, etc.
   Combining (H1)-(H6), these improve to 149 clusters with sizes 17, 16, 9, 8, 5, etc. 
   We note that (H1)-(H4) can explain the semantic behind certain asset transfers. Also, the link between master and phishing, which can be captured by APT-based heuristics, is not visible to traditional heuristics if the master uses a contract to attack and does not fund phishing directly.

       \begin{figure}[htb]
       \centering
       \includegraphics[width=1\linewidth]{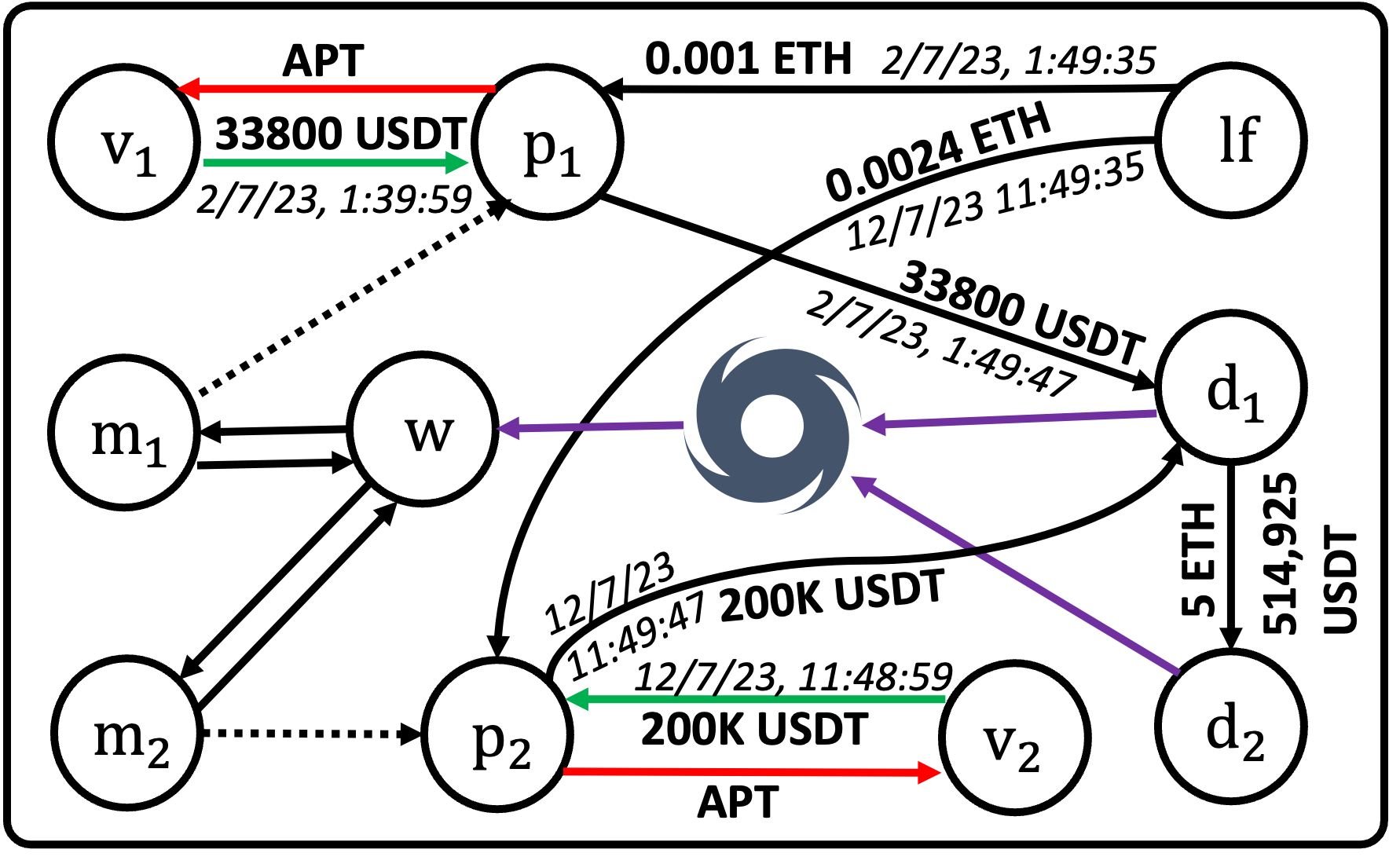}
       \caption{Examining two depositors $\don$, $\dt$, and one withdrawer $\w$, which share \textit{four} identical scam signatures. The scam cycle: \underline{\textbf{Step 1}} $\w$ withdrew 9.95 \ETH from TC and funded masters $\mo$ and $\mt$; \underline{\textbf{Step 2}} $\mo$ and $\mt$ coordinated APTs involving phishing addresses $\po$ and $\pt$; \underline{\textbf{Step 3}} $\po$ and $\pt$ received victim transfers; \underline{\textbf{Step 4}} $\po$ and $\pt$ received \ETH from the laundering funder $\lf$ to cover the fees and transferred the victim funds to depositor $\don$; \underline{\textbf{Step 5}} $\don$ deposited the scam proceeds back into TC, and also via the depositor $\dt$.}
       \label{fig:placeholder}
   \end{figure}

   \textbf{Example.} We consider the following two depositors ($\don$: \alinkLine{0xdaa6849270123f7c0b437cbd4ed24ca24825c838}, $\dt$: \alinkLine{0x536c3985d527b5ed823e6aa4f197e5bcf97ca53f}) and one withdrawer ($\w$: \alinkLine{0xe6ac0419c8737e6707d2d5c95c445acab860772e}).
   Being involved are also two phishing addresses ($\po$: \alinkLine{0xDA63e7BEEbFEeedB4CCfc7c295A54cEd3329FB95}, $\pt$: \alinkLine{0x9Fc67C82051D3C5839D4f1d35D9C37fcb58EaD98}), two masters ($\mo$: \alinkLine{0xA8f409e600AC692f88038465ca14917aDF5fD47B}, $\mt$: \alinkLine{0x675780CBd38AEEbE2Dc127F3676AA2aAC676FC3F}), and a laundering funder ($\lf$: \alinkLine{0x4a9B236E55A0dC226648018EF89B3e9FD41FF6fF}). 
   First, $\w$ withdrew 9.95 \ETH from TC 10, then funded several masters including $m_1$ and $m_2$. These masters coordinated thousands of token APTs, which involved phishing addresses such as $\po$ and $\pt$, before transferring the remaining amounts back to $\w$. Based on the event timestamps, we observe that once these phishing addresses received the transfers from victims, the (common) laundering funder $\lf$ immediately transferred some \ETH in so that they can pay fees to transfer the proceeds to $\don$, which deposited \ETH into TC. This depositor also transferred large amounts to $\dt$, which then also deposited into TC, ending a successful scam cycle.

    We end this section by adding a remark that apart from address grouping, we also discuss in Appendix~\ref{app:TC_grouping} a few intriguing case studies in which we attempted to use scam signatures to match potential TC depositors and withdrawers.

    \section{Related Work}
    \label{sec:literature}
    
    \textbf{APT Scams.} The rise of address poisoning scams has recently attracted attention from the security research community~\cite{YeHongZhangYang_WWW24,GuanLi_CCS24,Chen_etal_NDSS2025,tsuchiya_USENIX_2025}. Ye \et{}~\cite{YeHongZhangYang_WWW24} investigated zero-value APTs between July 2022 and June 2023 for ERC-20 tokens. Guan and Li~\cite{GuanLi_CCS24} covered a wider range of address-poisoning scams on Ethereum between 11/2022 
    and 2/2024 using \USDT and \USDC and their counterfeits, covering genuine-token APTs and fake-token APTs. Chen \et{}~\cite{Chen_etal_NDSS2025} studied the same types of address poisoning scams as~\cite{GuanLi_CCS24} between 12/2022 and 10/2023, but extended the token range to cover a few other high-value tokens on Ethereum such as \ETH, \WETH, \stETH, \WBTC, \BUSD, in addition to \USDT and \USDC. Tsuchiya~\et~\cite{tsuchiya_USENIX_2025} further included BEP-20 tokens on BNB Smart Chain in their investigation. 
    
    The main focus of prior work on APTs~\cite{YeHongZhangYang_WWW24,GuanLi_CCS24,Chen_etal_NDSS2025,tsuchiya_USENIX_2025} includes: (i) APT detection and statistical characterisation, (ii) victim and financial loss analysis, and (iii) address clustering using the scam-operation principle based on relationships among master addresses, phishing addresses, and APT/fake-token contracts. One exception is that Chen {\et}~\cite{Chen_etal_NDSS2025} performed address clustering based on the cashing-out/laundering process, however, ignored token swaps with DEXs. Tsuchiya {\et}~\cite{tsuchiya_USENIX_2025} also investigated the possibility of scammers using GPUs to generate phishing addresses that look similar to benign ones. We identify four research gaps in the state-of-the-art: (i) limited understanding of APT funding mechanisms, (ii) lack of formalisation and systematic analysis of behavioural scam signatures, (iii) insufficient investigation of how scam proceeds are laundered through public services (including CEXs, DEXs, and mixers), and (iv) the absence of studies on tracing activities across mixers such as Tornado Cash. We note that Tsuchiya {\et} was the first to observe that some large APT groups use similar APT and fake-token contracts~\cite{tsuchiya_USENIX_2025}.

    \textbf{Address grouping/clustering} is a fundamental problem in blockchain intelligence that has been the subject of extensive academic research (\cite{HarriganFretter_2016,MoserNarayanan_FC_2022,LinoyStakhanovaMatyukhina_CNSM_2019,Victor_FC_2020,Beres_etal_DAPPS_2021} and references therein) as well as industry commercial development (\cite{CTPro, Chainalysis_DataAccuracy_2024,TRM_Monero_2024}).   
    For law enforcement, identifying groups of addresses likely controlled by the same entity is valuable, as it can help link fraudulent activities (e.g. conducted by one address) to real-world identities (e.g. when another address in the group interacts with a KYC-compliant CEX). Most existing approaches rely on transaction histories, neighbourhood relationships, and address-level features.
    
    \textbf{Tornado Cash demixing} is a special case of address clustering. Matching depositors and withdrawers on TC is especially challenging due to its zero-knowledge-proof-based privacy mechanism, the lack of ground truths, and is often done case by case (see, e.g. ZachXBT's investigation~\cite{ZachXBT_Lazarus_TC}). State-of-the-art research is based on either heuristic approaches~\cite{Wu_et_Tutela_arxiv_2022, WangChaliasosQinZhouGaoBerrangLivshitsGervais_WWW_23} or machine learning~\cite{Du_et_al_TIFS24}.

    \section{Limitations}
    \label{sec:limitation}

    We acknowledge potential false positives/negatives in our datasets. 
    The APT dataset (Section~\ref{sec:APT_datasets}) may contain false positives where a user has addresses with identical first and last digits, either accidentally (this is rare according to~\cite{tsuchiya_USENIX_2025}) or intentionally. It may also miss APTs in which phishing and benign addresses match fewer than three first and last digits (should also be rare given that such attacks are less effective). It may miss genuine-token APTs that use tokens out of our top 100, which could have been popular before, but not anymore when we built our list. We can also miss fake-token APTs with symbols that do not conform to our definition. 
    
    The search for masters of each phishing address (Section~\ref{sec:master_distributor_datasets}) may fail for some rare cases, e.g. when the phishing address of an active nonzero genuine-token APT only has one coin funder and no token funder as anticipated. The reason is that the phishing address could have performed a swap of \ETH for the necessary token used in the APT. This is rare because of the costly economic: swapping for tokens requires multiple steps and is usually more expensive than directly transferring tokens (or bundled in a batch contract call). The master fundings and APTs may not be paired correctly from time to time (as transactions might not be perfectly ordered in time), leading to noisy data when gathering the master's signatures. However, we purposely design the consistency scores and flattened signatures to eliminate such noise. 

    Finally, our collected data do not cover the period before 11/2022 and after 5/2025, which could lead to edge cases due to insufficient data, especially for transactions occurring near the period boundary.
    
    \section{Conclusions and Future Work}
    \label{sec:conclusions}

    In this work, we study important aspects of Address-Poisoning Transfer scams on Ethereum that have previously been overlooked in the literature, including funding mechanisms, scam signatures, and proceeds laundering. Our proposal of scam signatures gives rise to a complementary approach to traditional address clustering methods, which often rely heavily on the asset transfer network among addresses. Restricting to a specific type of scam provides abundant context for clustering, as scammers often follow distinctive operational patterns. Using scam signatures allows for the potential tracing of scammers through mixers, such Tornado Cash. This method succeeds where traditional network-based approaches fail, specifically when there are no direct or indirect transfer activities between depositors and withdrawals and their nearby neighbours. Extensions of this approach to other types of on-chain scam such as Rug Pulls and Ponzi are left for future work.  

    \rev{We also observed that APT scammers sometimes reinvested their scam proceeds to initiate a new round of scams. We leave the investigation of scam reinvestment for future research.}

    \bibliographystyle{plainurl}
    \bibliography{APT_Network}

    \appendix
    \section{Open Science}

To support reproducibility and open science principles, we provide the following artefacts:

\begin{itemize}[leftmargin=*]
    \item \textbf{Codebase:} The source code for our data collection, clustering, and analysis is available at \url{https://anonymous.4open.science/status/apt-network}.
    \item \textbf{Dataset:} The anonymised datasets generated and used for evaluation are uploaded to the same Codebase repository
\end{itemize}

\subsection{Data Artefacts}
The submitted artefact contains a structured data snapshot intended to support schema inspection, validation of the construction logic on representative records, and reproduction of analyses over the released derived artefacts.
Unless otherwise specified, all paths are relative to the \texttt{dataset-submission/} directory. The artefacts herein are not intended to constitute a complete reproduction of all raw or intermediate datasets used in this work.

\subsubsection{Address-Poisoning Samples}
We provide a sample of the three APT datasets: the coin subset, the genuine-token subset, and the fake-token subset. These subsets are stored in the \texttt{normal/}, \texttt{token/}, and \texttt{fake-token/} subdirectories of \texttt{address-poisoning/}. They cover coin transfers, genuine-token transfers, and fake-token transfers.

\subsubsection{Victim Transfer and The Laundering Graph}
We provide the Victim transfer list, which contains information about the incidents, such as how much a victim has lost and which APT caused it. The laundering graph is provided in \texttt{laundering-dataset/}. Each row corresponds to a traced laundering edge from the algorithm \flf.

\subsubsection{Funding activities}
We provide multiple csv files that record funding operations: (1) from distributor to master; (2) from master to APT phishing; and (3) funding to successful phishing addresses before or during laundering activity. All under
\texttt{funding-activities/}.

\subsubsection{APT Scam Signatures}
We provide multiple csv files that list all the scam signatures of the master addresses of interest. Available under \texttt{scam-signatures/}.

\subsubsection{Tornado Cash Grouping}
We provide csv files that show identified connections of Tornado Cash users via different heuristics. Available under \texttt{tc-grouping/}.

\subsubsection{Market Data and Fake Tokens}
We provide the pricing data of genuine ERC20 tokens examined in this study; along with the fake-token addresses that mimic those genuine ERC20 in the path \texttt{resources/data}.

\subsubsection{Excluded data}
The complete database used is not redistributed because it contains full-scale public Ethereum transactions consisting of normal and internal transactions, token transfers, and receipt tables, whose size makes artefact hosting impractical. The artefact set also excludes private database configuration files, local caches, full PostgreSQL chain tables, and intermediate working files that support our internal analysis.

    \section{Ethical Considerations}

    This research involves the analysis of public blockchain transactions, specifically targeting APT phishing scams and subsequent fund laundering. The following ethical protocols govern the methodology and publication of findings:

\begin{itemize}[leftmargin=*]
    \item \textbf{Data Privacy and Acquisition:} All data analysed in this study consists of publicly available transaction logs from the blockchain ledger. Address attributions and identity heuristics utilise pre-existing public labels sourced from block explorers (e.g., Etherscan). No private, proprietary, or off-chain personally identifiable information (PII) was collected, nor was any independent active probing conducted to link addresses to real-world identities.

    \item \textbf{Address Clustering Limitations:} The address clustering technique based on scam signatures is designed to trace illicit APT scam proceeds and identify associated malicious actors. It targets addresses exhibiting specific scam-related behavioural signatures and is not intended to de-anonymise legitimate users of mixing services. In the absence of such signatures, legitimate users are unlikely to be clustered by our method. 
    
    \item \textbf{Dual-Use Assessment:} The formalisation of scam signatures and laundering patterns presents a theoretical dual-use risk, as malicious actors could analyse these heuristics to evade future detection. However, public disclosure of these mechanisms is necessary to advance defensive countermeasures, enable the development of robust detection models, and ultimately protect blockchain scam victims.
\end{itemize}

\section{Generative AI Usage}

GitHub Copilot, Cursor, Codex were utilised as coding assistants in the writing of the software implementation for this research. All AI-generated code was manually reviewed, executed, and validated by the authors to ensure functional correctness and accuracy prior to inclusion in the final experimental framework. This paper was edited for grammar and style using Writefull and ChatGPT.

    \section{Data Collection}
    \label{app:data_collection}

    \subsection{Fake Token Data Collection}
    
    To identify fake tokens, we design a matching algorithm to find these traits: (1) Fake token, $T$, has the same symbol as one of the top-100 tokens with a different contract address; (2) $T$ has a symbol of edit-distance 1 from a top-100 token symbol and either (2a) has a flawed contract permission logic or (2b) does not have a verified contract on Etherscan.
    
    \textbf{Step 1 - Data Collection.} We retrieve metadata for the top 600 cryptocurrency tokens from CoinGecko's public API by market capitalisation, establishing a baseline dataset of legitimate token symbols and their corresponding contract addresses. This is an attempt to address false positive cases, such that some symbols trying to mimic a token that is not in the top 100, but is highly similar (e.g. \texttt{oUSDT} is highly similar to \USDT, so  \texttt{oUSDT} will be classified as counterfeit). Subsequently, we obtained the complete list of unique contract addresses from the token transfer dataset from Google BigQuery to extract complete contract interaction data.

    \textbf{Step 2 - Token Analysis and Matching.} This step implements a dual-criterion detection system. \textbf{Step 2a - Type 1 detection.} We employ string similarity analysis similar to that of Taro \textit{et. al.}~\cite{tsuchiya_USENIX_2025}. This step uses three primary pattern recognition strategies to normalise a symbol:

\begin{enumerate}[leftmargin=*]
    \item Additional character insertion: where tokens append extra characters to legitimate symbols such as transforming \USDT to \texttt{USDTT}, 
    \item Leetspeak substitution recognises number-letter replacements such as converting \DAI to \texttt{D4I},
    and 
    \item Unicode homoglyph substitution identifies visually similar characters from different character sets, such as Cyrillic letters substituted for Latin equivalents. For example, by substituting the Latin `S' and `C' with the Cyrillic homoglyphs `S' (U+0405) and `C' (U+0421), the ticker appears identical to \USDC.
\end{enumerate}

    After normalisation, the edit distance of the symbol from its legitimate symbol is calculated. At this point, tokens that have an edit distance of 0 from the real token in the top 600 list are deemed fake. \textbf{Step 2b - Type 2 detection.} 
    If the token in step 1 has an edit distance of 1, we implement behavioural contract analysis, similar to the approach of Guan and Li~\cite{GuanLi_CCS24}. Specifically, the system attempts to trigger the \texttt{transfer} function using test accounts with insufficient balances and the \texttt{transferFrom} function without proper allowances. Legitimate contracts should reject these operations, while fraudulent contracts often implement permissive logic allowing unauthorised transfers, thereby revealing their malicious nature. We also check if the token address exists in the list of Etherscan verified contracts. Therefore, a token symbol of distance 1 is fake if its contract is flawed or not verified on Etherscan. In total we collected a list of 55,064 fake-token addresses.

    \subsection{Tornado Cash Data Collection}
    \label{app:tc_data_collection}

    \begin{table}[h]
        \centering
        \small
        \caption{The numbers of deposit and withdrawal transactions in Tornado Cash \ETH pools {\rev{(11/2022--05/2025)}}. These account for about {\red{98\%}} of all activities on Tornado Cash in this period.}
        \label{tab:TC}
        \begin{tabular}{
        @{}
        S[table-format=3.1] 
        S[table-format=6.0, group-separator={,}, group-minimum-digits=4] 
        S[table-format=6.0, group-separator={,}, group-minimum-digits=4]
        @{}}
            \toprule
            {Pool} & {No. deposits} & {No. withdrawals} \\
            \midrule
            100    & 7748           & 7025              \\
            10     & 10840          & 10446             \\
            1      & 10264          & 10045             \\
            0.1    & 6945           & 6796              \\
            \bottomrule
        \end{tabular}
    \end{table}

    \section{APT Datasets}
    \label{app:APT_datasets}

    \subsection{Dataset Construction}
    \label{rm:APT_Construction}        
        
    We include in this appendix the sub-procedures for finding coin APTs and fake-token APTs.
    
    \begin{algorithm}[htb!]
        \caption*{Coin APT Dataset Construction}
        \begin{algorithmic}
            \STATE\textbf{Step 1.} Identify the set $\tsc$ of all \textit{suspicious} normal transfers with (historical) values at most 3 USD 
            in which \textit{both} senders and receivers are EOAs and not identical (see Remark~\ref{rm:APT_Construction}). 

            \STATE\textbf{Step 2.} Based on $\tsc$, construct the set $\tac$ of \textit{coin} APTs as follows. For each transfer $t=(p,a)\in \tsc$, where $p$ and $a$ are potential \textit{phishing} and \textit{target} address, respectively, search the normal transaction history of $a$. If there is a \textit{nonzero} transfer of the same coin $\ell =(a,b)$ that happened before $t$ such that $b \neq p$ but $p$ and $b$ share at least three first and three last characters (ignoring the case), then include $t$ in $\tac$. Here $\ell$ is a \textit{legitimate transfer} associated with $t\in \tac$ and $b$ its \textit{benign} address.    
        \end{algorithmic}
    \end{algorithm}

    \begin{algorithm}[h]
        \caption*{Fake-Token DVT Dataset Construction}
        \begin{algorithmic}
            \STATE\textbf{Step 1.} Identify the set $\tsf$ of fake-token transfers in which \textit{both} senders and receivers are EOAs and not identical.
    
            \STATE\textbf{Step 2.} From the set $\tsf$ of fake-token transfers identified in Step 1, construct the set $\taf$ of \textit{fake-token} APTs and assign relevant labels to them as follows. For each transfer $t=(s, r) \in \tsf$ of a fake token $\tau$, first check if it is a \textit{forward} APT, by searching the transaction history of $r$. If there was a nonzero transfer of $\tau$'s genuine counterpart $\ell=(r,b)$ that happened before $t$ such that $b \neq s$ but $b$ and $s$ share at least three first and last digits, then include $t$ in $\taf$ as a \textit{forward} APT with $s$ as the phishing address. If $t$ doesn't satisfy that condition, then examine the transaction history of $s$ and look for a nonzero transfer $\ell = (s, b)$ of $\tau$'s genuine counterpart that happened before $t$ such that $b \neq r$ but $b$ and $r$ share at least three first and last digits (ignoring the case). In that case, include $t$ in $\taf$ as a \textit{backward} APT with $r$ as the phishing address. In either case, we refer to $\ell$ as the \textit{legitimate transfer} associated with $t$ and $b$ its \textit{benign} address. Lastly, if the sender of the normal transaction with the same hash as $t$ coincides with the phishing address then $t$ is labelled \textit{active} APT. Otherwise, it is \textit{passive}, and the sender of the normal transaction is referred to as the APT \textit{coordinator}. 
        \end{algorithmic}
    \end{algorithm}

    A few important points to clarify the APT dataset construction.

    In Step~1, we require that both senders and receivers of an APT must be EOAs. This is because while scammers can use contracts to perform multiple transfers simultaneously (for funding or scamming) to save gas cost, it is unlikely that the phishing address itself is a contract address for several reasons: 1) deploying a contract requires on-chain storage and computation cost, which is expensive, and 2) a single contract can only be used as a look-alike phishing address for very few target addresses (due to the randomness of addresses), which would further increase the cost of APT. 
    
    In Step~2, we require that the legitimate transfer $\ell$ has a nonzero value, for otherwise we may mistakenly identify a zero-value victim-to-phishing transfer as a legitimate one. Note that as opposed to Guan-Li~\cite{GuanLi_CCS24}, we do \textit{not} require that $\ell$ is non-suspicious in the first two procedures, because it is possible that $\ell$ is a test transfer of a small value and hence would belong to the set of suspicious transfers in Step~1. 

    Lastly, to speed up the search for APTs, we apply the method from Tsuchiya \et~\cite{tsuchiya_USENIX_2025} and let the algorithm scan for a potential legitimate transfer within a 20-minute window preceding the APT of interest, and only searching the entire history of the relevant address if such a transfer is not found.
    
\subsection{Analysis}
Our analysis encompasses 56,363,713 APTs spanning 918 days of blockchain activity (1 November 2022 -- 7 May 2025), providing insight into the evolution and sophistication of address poisoning attacks.

Our token APT datasets and those from Tsuchiya {\et}~\cite{tsuchiya_USENIX_2025} overlap significantly, with 16,076,923 common records (30.6\% of ours, 92.6\% of Tsuchiya et al.). The non-overlapping parts (36,431,096 unique to ours and 1,289,030 unique to Tsuchiya {\et} are due to different temporal coverage (ours 11/2022--05/2025 compared to 7/2022--6/2024); another reason is the differences in fake token identification - we follow the approach in Guan-Li~\cite{GuanLi_CCS24}, which focuses on fake tokens mimicking the top 100 tokens by market capitalisation. Table~\ref{tab:APT-dataset-comparison} summarises the transaction type data collected. 

\begin{table}[htb]
\centering\footnotesize
\caption{Comparison of our APT Dataset against relevant publications showing number of transactions grouped by coin, token, and fake token.}
\label{tab:APT-dataset-comparison}
\begin{tabular}{@{}
                l
                l
                S[table-format=9.0]
                S[table-format=3.1]
                S[table-format=9.0]
                S[table-format=2.1]
                l
                @{}}
                                            \toprule
                        &                   & \multicolumn{2}{c}{\textbf{Our Dataset (\%)}}        & \multicolumn{2}{c}{\textbf{CMU Dataset (\%)}}         & \textbf{PTXPhish}   \\ 
                                            \midrule
\multirow{3}{*}{\makebox[0pt][c]{\rotatebox[origin=c]{90}{\textbf{Coin}}}}          & Zero              & 667483  & 17.3              & {--}     &      & --  \\
                                            & Nonzero          & 3188211 & 82.7              & {--}     &      & --  \\ 
                                            \cmidrule(l){2-7}
                                            & \textbf{Total}    & 3855694 &                   &          &      &     \\
                                            \midrule
\multirow{7}{*}{\makebox[0pt][c]{\rotatebox[origin=c]{90}{\textbf{Genuine Token}}}} & Zero              & 12121499 & 81.9             & 7185298 & 95.9 & 104 \\
                                            & Nonzero          & 2678450  & 18.1             & 308881  & 4.1  & 22  \\
                                            & Forward          & 2910129  & 19.7             & 491757  & 6.5  & --  \\
                                            & Backward         & 11889820 & 80.3             & 7002422 & 93.5 & --  \\
                                            & Active           & 2448299  & 16.5             & {--}    &      & --  \\
                                            & Passive          & 12351650 & 83.5             & {--}    &      & --  \\ 
                                            \cmidrule(l){2-7}
                                            & \textbf{Total}    & 14799949 &                  & 7494179 &      & 126 \\
                                            \midrule
\multirow{5}{*}{\makebox[0pt][c]{\rotatebox[origin=c]{90}{\textbf{Fake Token}}}}    & Forward           & 102034   & 0.3              & 134233  & 1.4  & --  \\
                                            & Backward         & 37606036 & 99.7             & 9737542 & 98.6 & --  \\
                                            & Active           & 6        & 0.0              & {--}    &      & --  \\
                                            & Passive          & 37708064 & 100.0            & {--}    &      & --  \\
                                            \cmidrule(l){2-7}
                                            & \textbf{Total}    & 37708070 &                  & 9871775 &      & 100 \\
                                            \bottomrule
\end{tabular}%
\end{table}

The APT dataset comprises 3,855,694 direct Ether transfers, 14,799,949 genuine ERC-20 token transfer events, and 37,708,070 fake-token transfer events. Due to Ethereum's protocol, each native \ETH transaction generates a single transfer event; these transfers are classified as 100.0\% active and forward-directed because only private-key holders can initiate native \ETH transactions. Genuine-token transfers are predominantly passive (12,351,650 transfers, 83.5\%) and backward-directed (11,889,820 transfers, 80.3\%). These patterns emerge from ERC-20's \texttt{transferFrom} mechanism, where a token owner grants approval to a spender who then executes transfers from victim accounts to phishing addresses. Fake-token transfers show an even stronger pattern: 37,708,064 transfers are passive and only 6 are active, while 37,606,036 transfers (99.7\%) are backward-directed. Like genuine tokens, multiple fake-token transfer events can occur within a single contract transaction call; counterfeit contracts can also implement arbitrary transfer logic and emit events independently of victim consent and authorisation.

\subsubsection{Token Distribution}
Table~\ref{tab:token_distribution} shows the distribution between tokens across the APT datasets. \USDT dominates the transfer volume and stablecoins (\USDT, \USDC, and \DAI) account over 98.8\% of all APTs. 

\begin{table}[htb]
    \centering
    \footnotesize
    \caption{Distribution of APT by token for both genuine token and fake token data shows \USDT and \USDC make up the majority of all APT activity.}
    \label{tab:token_distribution}
    \begin{tabular}{
      @{}l
      S[table-format=8.0, group-separator={,}]
      S[table-format=2.1]
      l
      S[table-format=8.0, group-separator={,}]
      S[table-format=2.1]
      @{}
    }
    \toprule
    \multicolumn{3}{c}{\textbf{Genuine Token}}
      & \multicolumn{3}{c}{\textbf{Fake Token}} \\
    \cmidrule(r){1-3}\cmidrule(l){4-6}
    {Token} & {Transfers} & {(\%)}
      & {Fake Token} & {Transfers} & {(\%)} \\
    \cmidrule(r){1-3}\cmidrule(l){4-6}
    \USDT  & 10524432 & 71.1 & \texttt{fake\_USDT} & 25726615 & 68.2 \\
    \USDC  &  3818807 & 25.8 & \texttt{fake\_USDC} & 11010611 & 29.2 \\
    \DAI   &   212010 &  1.4 & \texttt{fake\_DAI}  &   454049 &  1.2 \\
    \WETH  &    45895 &  0.3 & \texttt{fake\_LINK} &   193640 &  0.5 \\
    \WBTC  &    36870 &  0.2 & \texttt{fake\_WBTC} &    93201 &  0.2 \\
    Others &   161935 &  1.1 & Others               &   229954 &  0.6 \\
    \cmidrule(r){1-3}\cmidrule(l){4-6}
    \textbf{Total} & 14799949 & 100.0
      & \textbf{Total} & 37708070 & 100.0 \\
    \bottomrule
    \end{tabular}
    \begin{tablenotes}[flushleft]
      \small
      \item \textit{Note:} Transfers are counted per token symbol across all address-poisoning
      transactions. Fake-token totals include both zero-value (0.2\%, 23{,}734) and
      non-zero-value transfers (99.8\%, 14{,}024{,}968).
      Stablecoins (\USDT, \USDC, \DAI) account for 98.8\% of APTs,
      consistent with prior work~\cite{tsuchiya_USENIX_2025,GuanLi_CCS24}.
    \end{tablenotes}
\end{table}

\subsubsection{Address Similarity}
Primary attack concentrations occur at low-complexity configurations, peaking at the 4-character head and 4-character tail (4,4) match configuration with 17.4 million attempts, followed by 15.9 million attempts at the (3,6) match configuration. A secondary cluster occurs at (7,8), with over 1.4 million attempts, indicating substantial computational resources used to generate these APT addresses.

\section{Tracking APT Profit Laundering}
\label{app:laundering}

In this appendix, we provide the algorithmic details of \dlf{}. For brevity, we present only the skeleton of our algorithm ignoring the swap logic and different token types.

    \begin{algorithm}[htb!]
    \label{algo:dynamic-laundering-flow}
    \caption*{\textbf{DynamicLaunderingFlow}$(V, T_V, D, k)$:}
        \begin{algorithmic}[1]
        \STATE{\textbf{Input}: victim-address set $V$, victim transfers $T_V$, TC pool addresses $D$ (or a set of public service addresses), depth limit $k$}
        \STATE $L \leftarrow T_V$ \algcomment{a dictionary: key = transfer ID, value = $(\texttt{src}, \texttt{dst}, \tau, \texttt{amt})$, where $\texttt{src}$ is sender, $\texttt{dst}$ is receiver, $\tau$ is time stamp,  $\texttt{amt}$ is laundering amount}
        \STATE $L_{\text{TC}} \la \varnothing$ \algcomment{laundering transfers entering Tornado Cash pools}
        \STATE $Q \leftarrow \varnothing$ \algcomment{the queue for nodes/addresses to be (re)processed}
        \STATE $\texttt{depth}[\cdot] \leftarrow -1$ \algcomment{initialise depth -1 for every node}
        
        \FORALL{$v \in V$}
            \STATE $\textsc{Enqueue}(Q,v)$; $\texttt{depth}[v] \leftarrow 0$
        \ENDFOR
        
        \WHILE{$Q \neq \varnothing$}
            \STATE $x \leftarrow \textsc{Dequeue}(Q)$
            \STATE $L_{\text{out}} \leftarrow \textsc{FindLaunderingFrom}(x,L,V,T_V)$ \algcomment{$L_{\text{out}}$ contains all laundering transfers from $x$}
            \FORALL{$\ell \in L_{\text{out}}$}
                \IF {($\texttt{ID}(\ell)\notin L$) or ($\texttt{ID}(\ell)\in L$ but with a smaller $\texttt{amt}$)}
                    \STATE $\textsc{Update}(L, \ell)$  \algcomment{if $\texttt{ID}(\ell)\notin L$: add $\ell$ to $L$, else: update $\texttt{amt}$}
                    \STATE $y \leftarrow \texttt{dst}$ \algcomment{the receiver of the laundering transfer $\ell$ from $x$}
                    \IF {$\texttt{depth}[y] = -1$}
                        \STATE $\texttt{depth}[y] \leftarrow \texttt{depth}[x]+1$ \algcomment{$y$ hasn't been visited before}
                    \ELSE
                        \STATE $\texttt{depth}[y] \leftarrow \min\{\texttt{depth}[y], \texttt{depth}[x]+1\}$
                    \ENDIF
                    \IF{$y \notin Q$ and $\texttt{depth}[y] < k$ and $y \notin D$}
                        \STATE $\textsc{Enqueue}(Q, y)$ \algcomment{unlike in standard BFS, $y$ is \textit{reprocessed} whenever there's a novel incoming laundering transfer}
                    \ENDIF
                    \IF{$y \in D$}
                        \STATE $\textsc{Update}(L_{\text{TC}}, \ell)$ \algcomment{similar to Line 13}
                    \ENDIF
                \ENDIF
            \ENDFOR
        \ENDWHILE
        \RETURN $L$, $L_{\text{TC}}$
        \end{algorithmic}
    \end{algorithm}
    
    \dlf~ takes as input the sets of victim addresses $V$, the set of victim-to-phishing transfers $T_V$, the set of Tornado Cash pool addresses $D$, and the depth limit $k$, and returns the dictionary $L$ of all predicted scam-proceed \textit{laundering transfers} within $d$ hops from the victim addresses, as well as the dictionary $L_{\text{TC}}$ of transfers that go into Tornado Cash. The algorithm first adds all victim addresses to the queue, $Q$, and sets their depth at $0$. While $Q$ is not empty, dequeue to obtain a node/address $x$ and call \flf() to obtain all laundering transfers $\Lo$ from $x$. For each $\ell\in \Lo$, if $\ell$ is new, i.e. either its ID is not already included in $L$, or its $\val$ is smaller than the new $\val$, update $L$ with $\ell$ (add $\ell$ to $L$ or update its $\val$). In either case, let $y$ be the receiver of $\ell$, setting its depth if necessary. If $y$ has not reached the max depth $k$ and is not a TC depositor, then add $y$ to the queue for processing. The algorithm always terminates because the number of nodes entering the queue (apart from the victim addresses) is bounded from above by the number of laundering edges.
    
    Unlike BFS, a node can be added to $Q$ and processed multiple times: whenever the algorithm identifies a novel incoming laundering transfer to a node $y$, it adds $y$ to the queue to recompute outgoing laundering transfers from $y$. This is crucial for accurately capturing the laundering transfers (see Example~\ref{ex:dynamic_laundering_flow} in Appendix~\ref{app:laundering}). 
    
    An output of \dlf~ is the total amount of all $\ell\in \LTC$ that gives an estimate of the total amount of scam proceeds that flow into Tornado Cash. Our algorithm operates under the assumption that scammers try to launder proceeds as fast as possible (discussed next). Furthermore, due to the depth limit, this estimation serves as a lower bound for the true amount.
    
    \begin{algorithm}[htb!]
    \label{alg:find-laundering}
    \caption*{\textbf{FindLaunderingFrom}$(x, L, V, T_V)$:}
    \begin{algorithmic}[1]
        \STATE{\textbf{Input}: an address $x$, the dictionary of laundering transfers $L$, the sets of victim addresses $V$ and victim transfers $T_V$
        \IF{$x \in V$}
            \STATE $\Lo \la $ victim transfers in $T_V$ from $x$
            \RETURN $\Lo$ \algcomment{if $x$ is a victim, return the (known) victim transfers from $v$}
        \ENDIF
        \STATE $\Li \leftarrow$ laundering transfers to $x$ in $L$, ordered by $\tau$ (time)}
        \STATE $\Lo \leftarrow \varnothing$ \algcomment{a dictionary of laundering transfers from $x$}
        \STATE $\To \la $ the list of out-going transfers from $x$ that occur after at least one transfer in $\Li$, ordered from old to new
        \FORALL{$t \in \To$ ordered by $\tau(t)$ from old to new}
            \STATE $\si \leftarrow \sum_{\ell \in \Li: \tau(\ell) < \tau(t)} \val(\ell)$ \algcomment{total incoming laundering by $\tau(t)$}
            \STATE $\so \leftarrow \sum_{\ell \in \Lo: \tau(\ell)<\tau(t)} \val(\ell)$ \algcomment{total outgoing laundering by $\tau(t)$}
            \STATE $a \leftarrow \min\{\si-\so, \val(t)\}$ \algcomment{$t$ launders as much as possible}
            \IF{$a>0$}
                \STATE $\lo \leftarrow \big(\texttt{key}=\texttt{ID}(t), \texttt{value}=(x,\text{dst}(t),\tau(t),a)\big)$
                \STATE Add $\lo$ to $\Lo$
            \ENDIF
        \ENDFOR
        \RETURN $L_{\text{out}}$
    \end{algorithmic}
    \end{algorithm}
    
    \flf~ takes as input an address $x$, the dictionary of currently known laundering transfers $L$, 
    and the set of victim transfers $T_V$, and outputs the dictionary $\Lo$ of outgoing laundering transfers from $x$. If $x$ is a victim, the algorithm simply returns the known victim transfers from $x$. Otherwise, let $\Li$ be the dictionary of laundering transfers in $L$ to $x$, and $\To$ be the ordered list of outgoing transfers from $x$ that occur after at least one transfer in $\Li$. For each $t \in \To$ ordered from oldest to newest, the algorithm computes the total amount $\si$ of incoming laundering transfers in $\Li$ and the total amount $\so$ of outgoing laundering transfers in $\Lo$ before the time $\tau(t)$. The balance of laundering funds, $\si-\so$, is the amount available to be transferred out at time $\tau(t)$ for $x$. The laundering amount $a$ for $t$ is thus the minimum of $\si-\so$ and $\val(t)$. If $a>0$, then the algorithm identifies $t$ as an outgoing laundering transfer $\lo$ from $x$ with $a$ as the laundering amount and adds it to $\Lo$. Our underlying assumption is that scammers prioritise laundering speed over other activities. In other words, if there are still proceeds not yet laundered, then the next outgoing transfer will use as much as possible. A step-by-step demonstration of how our algorithm works can be found in Example~\ref{ex:dynamic_laundering_flow} in Appendix~\ref{app:laundering}.

    We use an example to demonstrate key steps in the \dlf{} algorithm presented in Section~\ref{subsec:flow_to_TC}.
   \begin{example}
        \label{ex:dynamic_laundering_flow}
        To illustrate how \dlf{} works, consider an example with three victims $V=\{v_1, v_2, v_3\}$, two phishing addresses $P = \{p_1, p_2\}$, two intermediate addresses $q_1$, $q_2$, and one TC pool address $d$ (see Fig.~\ref{fig:dynamic_laundering_flow}). Set the exploration depth $k=3$. We have $T_V = \{t_1,t_2,t_5\}$, with values given in Fig.~\ref{fig:dynamic_laundering_flow}, e.g. $\val(t_1)=10$. The label $\val(\ell_i)$ in the figure refers to the laundering amount carried by the corresponding transfer $t_i$, which can be \textit{smaller} than $\val(t_i)$. We assume that $t_i$ occurs before $t_j$ if $i<j$.
        \begin{figure}
            \centering
            \includegraphics[width=1.0\linewidth]{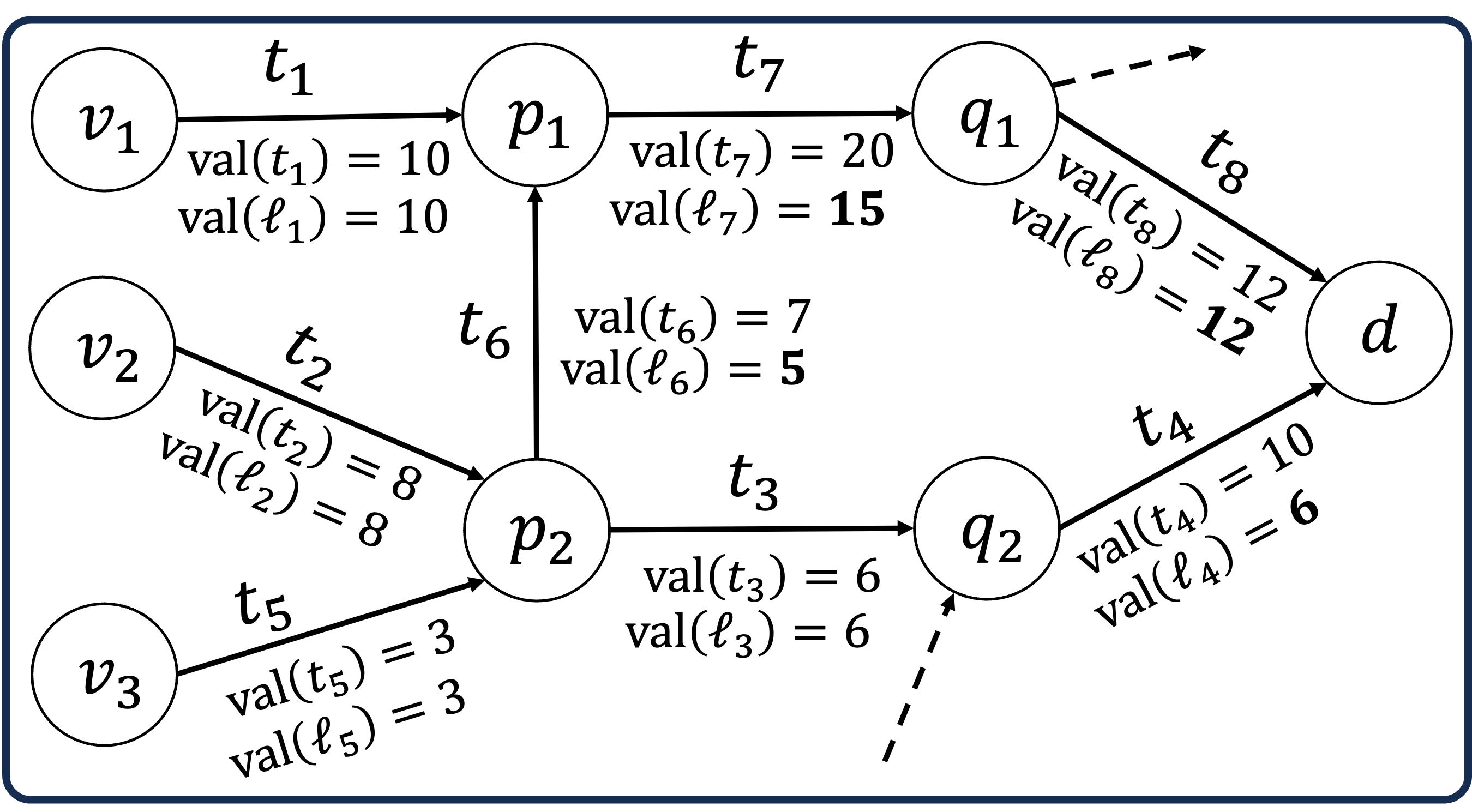}
            \caption{Illustration of \dlf. Here $v_i$'s, $p_i$'s, $q_i$'s, and $d$ represent victim, phishing, intermediate, and TC pool addresses, whereas $t_i$'s and $\ell_i$'s represent the transfers and the laundering transfers, respectively.}
            \label{fig:dynamic_laundering_flow}
        \end{figure}
        The queue $Q$ first contains $v_1,v_2,v_3$ (as of Line 6-7 in \dlf). After these victim addresses are processed, the dictionary of laundering transfers $L=\{\ell_1,\ell_2,\ell_3\}$, where $\ell_i$ is identical to $t_i$, $i=1,2,3$. Also, $Q = (p_1, p_2)$.
    
        In Iteration~1, $p_1$ is dequeued from $Q$, which has $\Li = \{\ell_1\equiv t_1\}$ with $\val(\ell_1)=\val(t_1)=10$. As $t_7$ occurs after $\ell_1$, $\Lo = \{\ell_7\}$ with $\val(\ell_7) = 10$ (it  will become 15 in a later iteration when $p_1$ is processed the second time). We have $L=\{\ell_1,\ell_2,\ell_3,\ell_7\}$, $Q = (p_2,q_1)$.
    
        In Iteration~2, $p_2$ is dequeued, which has $\Li = \{\ell_2,\ell_5\}$, $\To = \{t_3, t_6\}$. First, $\ell_2$ occurs before $t_3$, and so, $\val(\ell_3) = \min\{\val(\ell_2),\\ \val(t_3)\} = \min\{8,6\} = 6$. Next, both $\ell_2$ and $\ell_5$ occur before $t_6$, and so, $\val(\ell_6) = \min\{\val(\ell_2)+\val(\ell_5)-\val(\ell_3), \val(t_6)\} = \min\{8+3-6,7\} = 5$. We have $L=\{\ell_1,\ell_2,\ell_3,\ell_5,\ell_7\}$, $Q = \{q_1, q_2, p_1\}$. Note that $p_1$ is enqueued the second time.
    
        Fast forward to Iteration~5, $p_1$ is dequeued, and its $\Li = \{\ell_1,\ell_6\}$, where $\ell_6$ is the newly identified laundering transfer into $p_1$ in Iteration~2. Following Line 9-11 in \flf, $\val(\ell_7)$ is updated from 10 to $15 = \min\{10+5, 20\}$. A typical BFS would miss this update. Eventually, the fund flow entering $d$ is $18=12+6$.
    \end{example}

\section{Anatomy of the Laundering Networks}
\label{sec:anatomy_of_the_laundering_networks}

We analyse what happens after phishing addresses receive victim payments. Starting from the victim-to-phishing dataset, we trace any record of victim transfer to phishing address of above \$1,000, yielding 1,832 seed transfers, and follow them for 15 hops using \flf{}. The goal is to characterise how stolen funds are organised and moved after the initial theft. All USD values are computed from oracle prices at the time of each transfer.

\subsection{Flow of Assets}
Asset flows show a shift from stablecoin-heavy theft to \ETH-heavy exit.
At entry, victims mainly transfer stablecoins: \USDT accounts for 1,012 victim-to-phishing edges totaling \$61.60M, and \USDC for 394 edges totaling \$20.14M.
\ETH is still material, with 327 edges and \$12.21M in losses, while \DAI accounts for 28 edges totaling \$2.00M.
Aggregate losses are heavily influenced by four \WBTC transfers totaling \$68.47M, including one historically notable but atypical transfer worth \$68.46M.

At observed exits, \ETH dominates: it carries \$47.68M across 1,149 edges, with \WETH contributing another \$4.83M across 19 edges.
The contrast is especially clear for clusters that ultimately route funds to TC.
Their entry assets total \$34.72M across 263 victim edges, of which \USDT and \USDC account for 239 edges and 79.77\% of value.
Direct inflows to TC are almost entirely \ETH, comprising 422 edges totaling \$38.09M, alongside one \USDT edge worth approximately \$100.
In short, funds are often collected in stable-value assets and exited through \ETH.

\subsection{Services}

Tornado Cash is the dominant service used for laundering APT scam proceeds.
We provide detailed usage statistics for Tornado Cash ETH pools in Table~\ref{tab:laundering-tc-pools}. Here we use historical USD price for ETH.

\begin{table}[htb]
    \centering
    \small
    \caption{Tornado Cash usage statistics for APT laundering (11/2022--05/2025).}
    \label{tab:laundering-tc-pools}
    \begin{tabular}{@{}S[table-format=3.1]rrrr@{}}
    \toprule
    {Pool} & Deposits & \ETH & USD (\$M) & \% Share \\
    \midrule
    100 & 173 & 16,301.98 & 34.90 & 91.62 \\
    10  & 125 & 1,193.24  & 2.98  & 7.81 \\
    1   & 66  & 65.97     & 0.20  & 0.52 \\
    0.1 & 58  & 5.80      & 0.02  & 0.05 \\
    \midrule
    {Total} & 422 & 17,566.99 & 38.09 & 100.00 \\
    \bottomrule
    \end{tabular}
\end{table}

\subsection{Laundering Cluster Case-Studies}
\label{app:laundering_cluster_case_studies}
\begin{figure*}[p]
    \centering
    \begin{subfigure}{\linewidth}
        \centering
        \includegraphics[width=\linewidth]{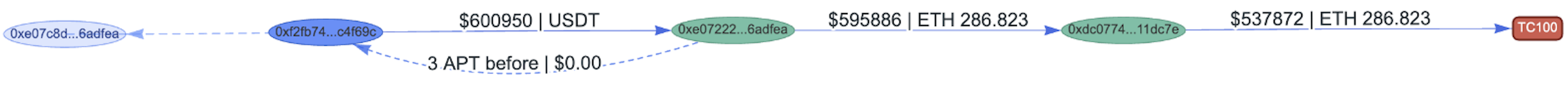}
        \caption{Minimal victim $\rightarrow$ phishing (\alinkLine{0xe07222cb87725f1e88ef48dd17e9a8ccb86adfea}) $\rightarrow$ depositor $\rightarrow$ Tornado Cash chain. The scammer tops up by 13.4 \ETH (see the history of \alinkLine{0xdc07748cbc104005bd077a4dd758440fc011dc7e}) to send 300 \ETH to the \texttt{TC-100} pool.}
        \label{fig:Cluster_15}
    \end{subfigure}

    \begin{subfigure}{\linewidth}
        \centering
        \includegraphics[width=\linewidth]{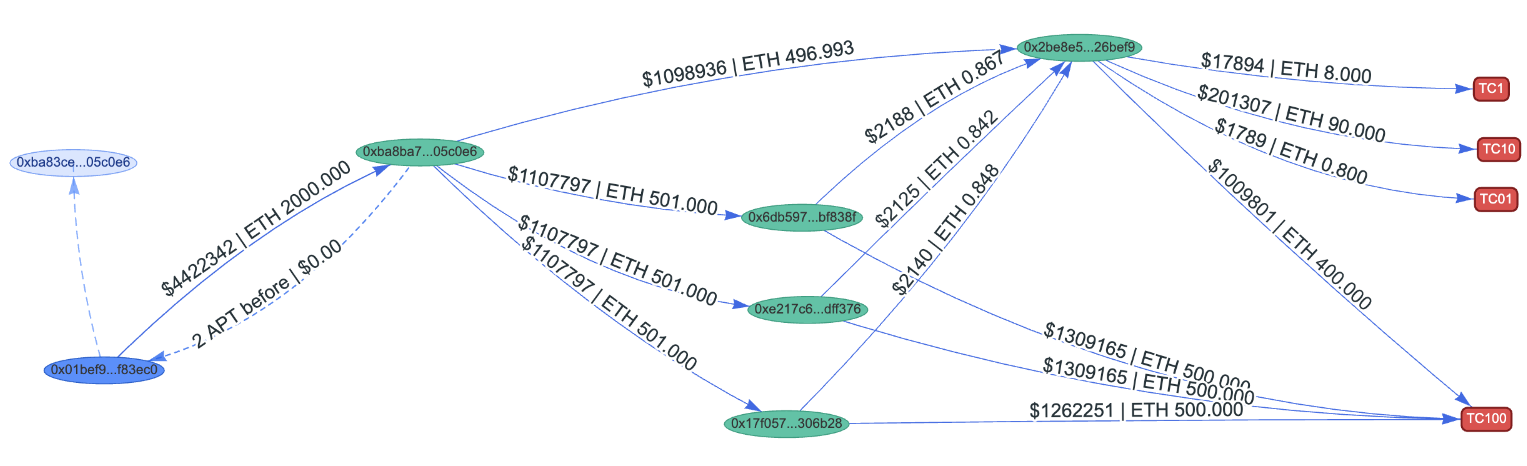}
        \caption{The phishing address (\alinkLine{0xba8ba758357d82a2862e1369d51c983a2a05c0e6}) successfully acquires 2000 \ETH, splits it into 4 sibling branches and consolidates to ensure a round number eligible for the \texttt{TC-100} pool. A total of 1998.8 \ETH is transferred to TC.}
        \label{fig:Cluster_2}
    \end{subfigure}
    
    
    \begin{subfigure}{\linewidth}
        \centering
        \includegraphics[width=\linewidth]{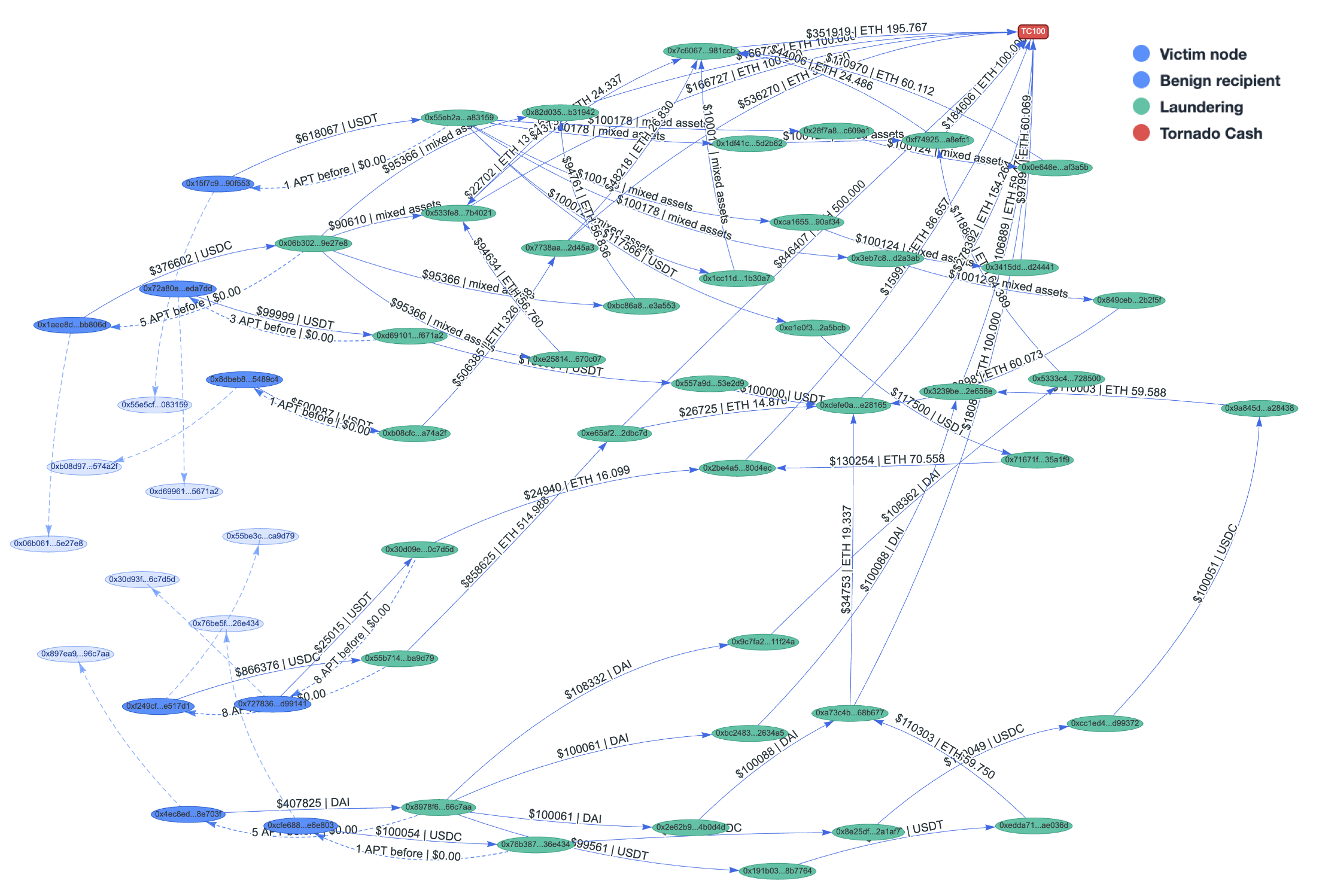}
        \caption{A large-graph case indicating an ongoing multi-target poisoning campaign. The internal flow is mixed-asset; the final cash-out converges on the \texttt{TC-100} pool (top). Some of the phishing addresses includes: \alinkLine{0x06b3022f191a957e169e9b4ce8913b14859e27e8}, \alinkLine{0xb08cfcbe73d3813648ca7f96e6178d7208a74a2f}, \alinkLine{0x55b714f59f71b65cc63201fc9fc8a817adba9d79} and more.}
        \label{fig:Cluster_4}
    \end{subfigure}
    
    \caption{Comparative analysis of asset flows across three exemplar clusters. Victim nodes are in blue, internal laundering nodes in green, and transfers out to TC in brown.}
    \label{fig:Combined-Clusters}
\end{figure*}

We present three exemplar clusters in Figure~\ref{fig:Combined-Clusters} to demonstrate APT asset flows towards transfers out to Tornado Cash.

Figure \ref{fig:Cluster_15} shows the simplest laundering topology among the three case studies. The phishing address appears three times in the victim's history prior to the theft. The stolen value then follows a minimal victim $\rightarrow$ phishing $\rightarrow$ depositor $\rightarrow$ Tornado Cash chain: \USDT is converted into 286.823 \ETH and forwarded to the final depositor. Manual inspection indicates that the scammer then adds their own 13.4 \ETH \txlink{0x2ad464aaa5d36000e2466e6fc934c605df9cb7531e9e6528d80c19e3eeaffd9a} from FixedFloat to round up to 300 \ETH to fit into the \texttt{TC-100} pool. This case also motivates us to design \flf{} to filter and constrain the laundering flow to avoid commingling with external funds.

Figure \ref{fig:Cluster_2} shows that the phishing address attacked twice before the 2000 \ETH theft. After receipt, they split the funds into four sibling branches. One branch, \texttt{0x2be8e5\ldots26bef9}, first receives 496.99 \ETH directly and later absorbs 2.56 \ETH in residual transfers from its siblings before sending 498.8 \ETH to Tornado Cash. This residue re-convergence inspired the node revisitation mechanism in \flf, rather than regular BFS.

Figure \ref{fig:Cluster_4} provides the complementary large-graph case. Several victim-facing branches are preceded by repeated APTs, indicating an ongoing multi-target poisoning campaign rather than a single incident. The internal flow remains mixed-asset, with multiple local consolidation points. The final cash-out converges on the \texttt{TC-100} pool.

    \section{Master and Distributor Datasets}
    \label{app:master_distributor_datasets}

    \subsection{Examples of Master-Phishing Interactions}
    \label{app:examples_master_phishing}

    \begin{example}[Coin APT, direct funding]
        \label{ex:coinAPT_direct_funding} \textbf{Fake\_Phishing327990}, which successfully got 72 million worth of \WBTC after the APT attack (see Example~\ref{ex:68_0_DVT}), received a direct funding of 0.00021 \ETH from the master \textbf{0xDCdd}\alink{0xDCddc9287e59B5DF08d17148a078bD181313EAcC} labeled \textbf{Fake\_Phishing343305}, which funded more than \textit{twenty thousand} phishing addresses. 
    \end{example}

    \begin{example}[Coin APT, contract funding]
        \label{ex:cointAPT_contract_funding} The master \textbf{0x1552}\alink{0x1552dE3db8B4c9E0f7e9a426623b4469D1f10931}, labeled \textbf{Fake\_Phishing463645}, called the same contract labeled \textbf{Fake\_Phishing464645}\alink{0xf396998cd79369fE3b4F336Ab2f5F438f238Ab2a} almost \textit{six thousand} times, each call funding multiple phishing addresses in \ETH. For instance, a single call on 12/3/2025\txlink{0xa136f83785440a3c16dc0caf35258408246f64de1bf6c3358f65dbaae2afb51b} funded 14 phishing addresses, most of which performed an APT worth 0.000001 \ETH with fee 0.00001155 \ETH, summing up exactly to the funding 0.00001255 amount \ETH. 
        Note that some of the phishing addresses, e.g. \textbf{0x2660}\alink{0x266098D6fE5F706516Cc24Eefd83E3321585A68D} also performed genuine-token APTs (e.g. \USDT/\USDC) and was funded by other contracts and masters.
    \end{example}

        \begin{example}[Master funding \ETH and \USDC directly]
        \label{ex:master_direct_funding_ETH_USDC} The master \textbf{0x3C6a}\alink{0x3C6a325bE174845bC4889C37be70B18DB4EFC2f1}, labeled \textbf{Fake\_Phishing442895}, funded more than \textit{ten thousand} phishing addresses using the same mechanism: first transferred \ETH,   then the token, which allows the phishing address to perform a token APT. For instance, it first transferred 0.00088027 \ETH\txlink{0x435a1cdbd991f7d190de791a0aebd1ab4d810d1ce4b9161a394d40eea82564d4} and then 0.2 \USDC\txlink{0xb66d9465d0d2b36ab400d31aebbb6b7a08036f2446ea15f6a650483c556e3ee2} twelve seconds later (1 block later) to the phishing address \textbf{0x0E0A}\alink{0x0E0A49391a6f9309cf6f576B470c7bBc1ce070F9}, which performed an APT of exactly 0.2 USDC in less than three minutes (11 blocks) later, with a fee of 0.00054222 \ETH (less than what it received). 
    \end{example}

    \begin{example}[Genuine-token APT, one coordinator, multiple coin funders]
        \label{ex:genAPT_multiple_coin_funders}
        The phishing address \textbf{0x2aFa}\alink{0x2aFa81f87BC9ABC0fFfE88cd92b8CBF5D4B38BCE} (labelled \textbf{Fake\_Phishing-\allowbreak 55731 on Etherscan)} received five coin-funding transfers from four different masters before it could cover the fee for the transaction\txlink{0x602406479e8050b6bf2b84c321549d908c7e1de6e3dce9bbc02c76ba26542672} that approved the contract labelled \textbf{Fake\_Phishing11700} to spend on its behalf (see Fig.~\ref{fig:gentokenAPT_coordinator}). 
        About six minutes (29 blocks) later, the master (coordinator) \textbf{0xefb3} (labelled \textbf{Fake\_Phishing11703} on Etherscan) called a function on this smart contract\txlink{0x3a927afd654d7c74b3420b9f85a0aba6ee85471da057895df54e939e16424d38} to perform 32 token transfers, corresponding to 16 APTs. In particular, the contract transferred 0.05 \USDT to the phishing address in one transfer, then in the next, it transferred 0.05 \USDT out of that node to a target node \textbf{0xD717}. 
        The same thing went for phishing addresses \textbf{0xfB68}\alink{0xfB68e1531B752aA77Fb1db8e199F934595264954} and \textbf{0x211C}\alink{0x211C50FedA71148a072aAB5741369228dFe2c882}. This was perhaps caused by the fluctuation of the gas price, impacting the scammer's prediction of the fee. 
    \end{example}

    \begin{example}[Genuine-token APT, one coordinator, no coin funders]
        \label{ex:genAPT_no_coin_funders}
        Consider the same phishing address \textbf{0x2aFa} as in Example~\ref{ex:genAPT_multiple_coin_funders}. It had no coin fundings before the master (coordinator) \textbf{0xc16a} (labelled \textbf{Fake\_Phishing48029} on Etherscan) called a function on the smart contract labelled \textbf{Fake\_Phishing48030}\txlink{0x4a25cdebe1a39e672b452e9c4628be56d3e21f42568b59245456f42a6833afac} to perform 80 (zero-value) backward APTs, including one involving this phishing node. In total, \textbf{0x2aFa} was involved in 14 such backward passive genuine-token APTs.  
    \end{example}

        \begin{example}[Successful fake-token APT, one coordinator] \label{ex:successful_fake_token_APT}
        The phishing address \textbf{0x4e7C1d6}\alink{0x4e7C1d68eD7bD6754e7ddD3863836b5cAec7ea87} were involved in two backward fake-token APTs, coordinated by the same master \textbf{0xA571}\alink{0xA57188bD032cf7fA7319608f0FD9F155b387205D} using the same scam contract\alink{0x4b018B99d990825079C9794c382DfAa8bbd395a6}. The APT\txlink{0xb13375c675d4fd021586aea0de717b104b982d86ec2ed0cb857aa94be32484af}, in which the phishing address received 120,763.518627 fake \USDC (identical to the amount of \USDC in the benign transaction\txlink{0xe1518167d0c6f7e244e94c9d540651a4a0791f47d6a372d749f5d5b71d98145d}) from the victim, was successful. The victim mistakenly transferred over 130,000 \USDC\txlink{0x05e62526912bf4dd6322929f28642f879a9316cc7a724327cc3ec49997dd4876} to the phishing address. Subsequently, the address \textbf{0x55e6}\alink{0x55e6e031c37B270aF3ebb788938bD32F9E6Cc19d} transferred 0.02 \ETH to the phishing address to cover the fees for exchanging USDC for \ETH on Uniswap, and for transferring the scam revenue of about 56 \ETH to another address. 
    \end{example}

    \begin{example}[Successful fake-token APT gaining 2000 \ETH] \label{ex:2000ETH}
    The fake-token APT \textbf{0x57d6}\txlink{0x57df61872793d02a33e27e3bb77ffb7f525a91f681183b5c72b3c15a089526c5}, which mimicked the benign transfer \textbf{0x69ea}\txlink{0x69ea28b1d395bc23e5a59641100c7b88c46a8cf601c132e535830ac34208c057}, led to a sizable victim transfer of 2000 \ETH\txlink{0xa8d0ac9c1f495af05b748749833e2b94012410132ac40c3aa0aa5200a65a6523} (worth 4.4 million then). See also Fig.~\ref{fig:Combined-Clusters}b in Appendix~\ref{app:laundering_cluster_case_studies} for the laundering.
    \end{example}

    \begin{example}
    \label{ex:funding_using_victim_transfers}
        The phishing address \textbf{0x06CF}\alink{0x06CFbd19BB436D8d75499d5cbb995C38018c7322} received the first funding transfer of 0.005 \USDC from the master \textbf{0xbb94} and then attacked the victim \textbf{0x81A8} with 0.005 \USDC, mimicking a benign transfer\txlink{0xb442a6c219f4bdb993bdad6b16ff69bb8ba7f74f1ad26731481a8f4fe6a29363}. The victim mistakenly transferred 5,000 \USDC to the phishing, which took 0.015 \USDC out of this transfer to send another APT to the same victim, before transferring the remainder (4,999.985 \USDC to another address). The victim fell again for this APT and lost another 5,000 \USDC. The phishing repeated the action, and succeeded the third time! It used the victim transfers to fund for three other APTs to \textbf{0x81A8} three more times before transferring all the remaining \USDC to another address.
    \end{example}

    \begin{example}\label{ex:funding_using_victim_transfers_ETH}
        The phishing address \textbf{0x3bF2}\alink{0x3bF2dE59432Ce0C257bfe3ed6A0375e840331737} sent nine consecutive coin APTs of 0.000001 \ETH, in which the first one mimicking a previous benign transfer\txlink{0x7951ffb45feb2420b9c07b3d7544ccb9651f1ebbc20f8c66dfb1b46972ab6ce1}. These APTs were funded with exact amounts (covering precisely the APT values and the transaction fees) by nine internal transactions. But once received a victim transfer of 19.2754893 \ETH, it used this to fund the next APT to the same victim before transferring the remaining amount (including some small APT amounts coming in from other scammers) to another address.
    \end{example}

    \begin{example} \label{ex:funding_after_successful_APT} 
    A zero-UDST APT~\txlink{0x6068f5fa7703f1dea289ca82c5f04e44e1e9602738adead78443bbaaf2f595a2} (and perhaps also the ones preceeding it) led to a victim transfer\txlink{0x3d22e9cb448498a32923f17badd8739146e020255e7f5a8eeda1c355c2d1ce3a} of 5,000 \USDT to the phishing address \textbf{0xCFab}\alink{0xCFabEF41fC0076f9736EDe647a39468A426667cA}. This phishing address received a funding of 0.5 \ETH\txlink{0x39b01f641f6ebc880cc031f6e9d827bbdba5a6cdee2c572cf0443f8e12cc28fe} from a laundering funder, which was subsequently used to cover the transaction fee for a swap of \USDT into \ETH via Metamask Swap. \textbf{0xCFab} then turned itself into a master (coordinator) by creating an APT contract and coordinated near 800 transactions involving 0-\USDT APTs. The cost of running such APTs was also covered by 10.33 \ETH transferred from \textbf{0x04C3}\alink{0x04C37956972727fc0C52aA4CD64aed18b0a62aEC}, another successful phishing node. Eventually, \textbf{0xCFab} transferred 6.1 \ETH and 0.083294513 \ETH to two aggregator nodes \textbf{0x0Eb9}\alink{0x0Eb9b84c20538A742022B2cc8C9995D0e9052322} and \textbf{0x51fb}\alink{0x51fb0E9F117A9be64FD9a3fd0daDf6d1Ae90FFac}, respectively. \textbf{0x0Eb9} deposited 40 \ETH into TC, whereas \textbf{0x51fb} continued to transfer more than 82 \ETH to another node, which then deposited 100 \ETH into TC. 
    \end{example}

    \begin{example}
        The successful phishing address \textbf{0x04C3}\alink{0x04C37956972727fc0C52aA4CD64aed18b0a62aEC} also behaved in the same way as the phishing address \textbf{0xCFab} mentioned in Example~\ref{ex:funding_after_successful_APT}. The master that funded this phishing address after it succeeded used mostly 3 Gwei as Max Priority but also 1.5 Gwei in one transaction (why the inconsistency?). Its three contract creation transactions all used 6 Gwei as Max Priority.
    \end{example}

    \subsection{Identifying the APT Distributors and Their Funders}
    \label{app:distributor_dataset}

    \textbf{Distributors.} We define APT \textit{distributor} addresses (or distributors for short) as those that funded the APT master addresses. More specifically, funding transfers for a master from its corresponding distributors can be identified by collecting all nonzero-value non-APT non-self-transfer incoming \ETH or genuine token transfers before its last APT-funding/coordinating transaction and extracting their EOA callers. We found 500,866 distributor transfers initiated by 420,997 unique distributors to support 7,508 master addresses. Non-service distributor addresses that funded the most number of masters include \alinkLine{0x38401e2af98ee72bc8bffeea2f395fc4d256d235} (funded 65 masters) and \alinkLine{0x0c7c3643de2f39ffd2dcc7d9d88a497f2ce1baa4}, which consistently used 1inch to fund 55 masters.   

    We identified 88 public service addresses that funded APT masters, led by FixedFloat (funded 345 masters), ChangeNOW (132), WhiteBIT (39), and Kraken (39). Note that these services implement KYC at various levels: Kraken and WhiteBIT require identity verification, while FixedFloat and ChangeNOW are non-custodial and do not require KYC.

    The distributor dataset can be used to determine master clusters (see Section~\ref{subsec:sig_contract_calls}) in which master addresses not only have similar scam signatures, but also fund each other. Such addresses are labelled as both masters and distributors. Overall, there are 4,428 distributors that are also masters that supported at least two APT-funding/coordinating transactions and funded other masters. There are 272,199 distributors that are also phishing addresses, which include a) phishing addresses that transferred the remaining amounts after receiving some funding back to masters, e.g. 
    \alinkLine{0xf0c3ce7e061816d545bffcb809d4a423ef3a92c7}, and b) successful phishing addresses that transferred to master addresses for either laundering or initiating new rounds of APTs (see Fig.~\ref{fig:example_DVT_network_TC}).

    \textbf{Distributor Funders.} Similarly, we define \textit{distributor funders} as addresses that transferred ETH to distributors before their last funding transfers to masters. An example of a distributor funder is \alinkLine{0xf4C4263ED3Abb1431c40929BF55321e843B17A65}. After receiving 9.95315 \ETH from Tornado Cash, it transferred about 5 \ETH to the distributor \alinkLine{0xF8Bc7b2C5985b7129125AcD87f78e09349e78206}, which subsequently funded dozens of masters via 1inch's Aggregation Router V5. As mentioned in Section~\ref{subsec:master_dataset}, distributors and distributor funders are intermediate addresses on the `tainted' ETH-transfer paths that link scam-related TC depositors and withdrawers to APT masters. When creating the tainted paths, we ignore distributors and distributor funders that are public-service addresses or TC relayer addresses to avoid obvious false positives. 

    We captured 656,762 unique distributor funders in our datasets (restricted to the set of 7,508 masters of interest). Among these, 138,009 are also distributors, 4,612 are also masters, 120,814 are also phishing addresses, and 994 are public-service addresses.
    
    \section{APT Scam Signatures}
    \label{app:scam_signatures}

    \subsection{Consistency Scores}
    \label{app:consistency_scores}
        Let $M=\{(v_i,f_i)\}_{i=1}^k$ denote a multiset of $k$ observed elements, where each distinct element $v_i$ appears with frequency $f_i$, $N = \sum_{i=1}^k f_i$ is the total number of observations, and $p_i = f_i/N$ is the relative frequency of the $v_i$. Then the consistency score is defined as $C(M)\triangleq 1-\log \ke / \log N$ if $N>1$, and $C(M) \triangleq 1$ if $N=1$, where $\ke$ is the \textit{effective} number of elements, defined as follows. Using the Inverse-Simpson index, we have $\ke = \ke(\textsf{IS}) \triangleq 1/\sum_{i=1}^k p_i^2$. If the Shannon entropy is used, then $\ke = \ke(\textsf{SH}) \triangleq e^{-\sum_{i=1}^k p_i\log p_i}$. 
        
        For instance, for $M=\{(1,99),(2,1)\}$, we have $\ke(\textsf{IS})(M)=1.02$ and $\ke(\textsf{SH})(M)=1.06$, representing the fact that although two values appear, most of the time it is just one.
        We use $C_{\textsf{IS}}(M)$ and $C_{\textsf{SH}}(M)$ to denote the corresponding consistency scores. 
    
        Note that $C(M)\in [0,1]$, and the closer it is to one, the more consistent the values of $M$ are, i.e. fewer values with larger frequencies. Clearly, $\cis(M)=\csh(M)=1$ if $M=\{(v_1,f_1)\}$ (only one element appears), and $\cis(M)=\csh(M)=0$ if $M=\{(v_1,1), (v_2,1),\ldots,\allowbreak(v_k,1)\}$ (each element appears exactly once). Note that the Simpson index and the Shannon capacity measures diversity of values, ignoring the total number of elements. For example, they give the same values for $M_1=\{(v_1,2), (v_2,1)\}$ and $M_2=\{(v_1,2000), (v_2,1000)\}$ as $p_i$'s are the same for both multisets. Our consistency scores, however, take into account the total number of observations $N$, and give much higher scores for $M_2$ ($\csh(M_2)=0.87 > \cis(M_1)=0.25$), which fits our purpose.

    \subsection{Guilt-by-Association) Graph and Signature Cohesion Score}
    \label{app:cohesion_score}

    We provide in this appendix the formal definitions of the guilt-by-association graph and the signature cohesion score.
    
       \begin{definition}[Guilt-by-Association Graph] 
        \label{def:guilt_by_association_graph}
        Given a set $M$ and $D$ of master and \textit{non-service} distributor addresses, respectively (see Section~\ref{sec:master_distributor_datasets}), the \textit{guilt-by-association (GBA) graph} $G(M,D) = (V,E)$ is a directed graph constructed via the following steps: Step 1) set $V = M$, Step 2) add the edge $e=(m_1,m_2)$ to $E$ if two masters $m_1\neq m_2$ in $M$ support the same phishing address, Step 3) for two masters $m_1$ and $m_2$ in $M$, if $m_1$ is a distributor of $m_2$ then add $e=(m_1,m_2)$ to $E$, Step 4) for each distributor $d \in D\setminus V$, if $d$ fund $h\geq 2$ masters $m_1,\ldots,m_h$ in $M$, then add $d$ to $V$ and $e_i=(d,m_i)$ to $E$ for $i=1,\ldots,h$.
    \end{definition}

        Note that existing work already used a version of Step 2 to group master/phishing addresses when two master addresses coordinate APTs involving the same phishing address~\cite{GuanLi_CCS24, tsuchiya_USENIX_2025}. However, they missed master-to-master and distributor-to-master \ETH transactions. Steps 3 and 4 are introduced in our work to fill that gap.

    \begin{definition}[Signature-Level Cohesion] 
        Let $M$ and $D$ be sets of master and distributor addresses, respectively, and $\sigma$ be a scam signature. Let $\sigma(m)$ denote the corresponding signature value of $m\in M$. Let $\{\sigma_1,\ldots,\sigma_r\}$ be the set of all distinct values of $\sigma(m)$, $m\in M$, so that each set $M_i \triangleq \{m\in M\colon \sigma(m) = \sigma_i\}$, $i=1,\ldots,r$, has at least two elements. Let $\{C_{i,j}\}_{j=1}^{k_i}$ be the weakly connected components (WCC) of the scam-operation graph $G(M_i,D)$ (see Definition~\ref{def:guilt_by_association_graph}). We define the cohesion score of the address group $M_i$ as 
        \[\coh_{M,D}(M_i)\triangleq \sum_{j=1}^{k_i}\binom{|C_{i,j}|}{2} / \binom{|M_i|}{2} \in [0,1].
        \]
        The cohesion score of the signature $\sigma$ is defined as
        \[
        \coh_{M,D}(\sigma) \triangleq \sum_{i=1}^r\sum_{j=1}^{k_i}\binom{|C_{i,j}|}{2}/ \sum_{i=1}^r\binom{|M_i|}{2}  \in [0,1].
        \]
    \end{definition}

    \subsection{Explaining the Copying Bot Case from~\cite{tsuchiya_USENIX_2025}}
    \label{app:copying_bot}

    The presumed copying bot \textbf{0x7D57}\alink{0x7D575a7C732D1c502c07f18BB822D29CF7DBf9E8} called two contracts to perform APTs using two type-0 transactions with gasLimit 165,807 and 53,082, respectively. On the other hand, the first contract was created by the master \textbf{0x9376}\alink{0x9376C9c73D8C7AC9b6ae159ADaA06b42dbBef1B4}, who repeatedly called it 314 times using type-2 transactions with gasLimit 18500000, maxFeePerGas 200 Gwei, and maxPriorityFeePerGas 3 Gwei. It turns out that this address belongs to \rev{Group CG1 of 1,699} master addresses (see Section~\ref{subsec:sig_contract_calls}), which all generated contracts with identical bytecodes and had consistent gas signatures. The second contract was created by another master \textbf{0x7DC4}\alink{0x7DC44D10499EE4c3F4bbbfCD6Ed428dDC7b99451}, which belongs to \rev{Group CG3} of 254 master addresses (see Section~\ref{subsec:sig_contract_calls}), which generated contracts with two distinct bytecodes and almost always used type-2 transactions and maxPriorityFeePerGas 1.5 Gwei. Thus, the copying bot, \rev{Group CG1, and Group CG3} have completely different contract and gas signatures, substantiating the observation in~\cite{tsuchiya_USENIX_2025}.

    \section{Grouping User Addresses on Tornado Cash via Scam Signatures}
    \label{app:TC_grouping}

    In this appendix, we explore an approach that could suggest pairs of matching TC deposit and withdrawal. More specifically, we first identify tainted TC depositors and withdrawers, which are associated with APT masters via tainted paths. Scam signatures from those masters are then collected and aggregated for such tainted TC user addresses. Finally, deposits and withdrawals that are closest in time and initiated by tainted depositors and withdrawers with similar scam signatures are suggested as potential candidate matches.
        
    \begin{figure}[htb!]
        \centering
        \includegraphics[width=1\linewidth]{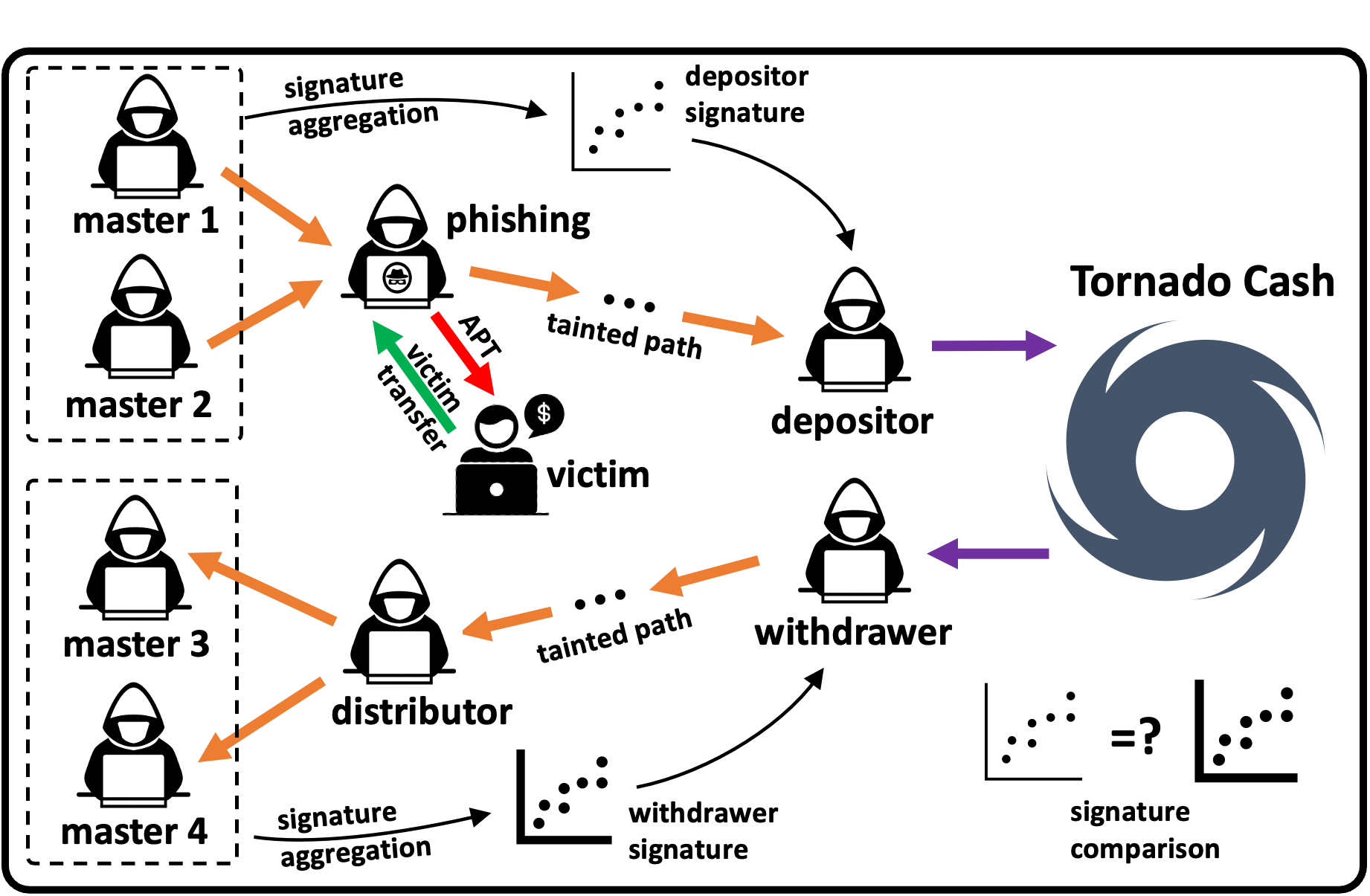}
        \caption{TC depositors and withdrawers are assigned scam signatures from their associated masters via `tainted' paths.}
        \label{fig:TC_grouping_approach}
    \end{figure}

    \begin{figure*}
        \centering
        \includegraphics[width=1\linewidth]{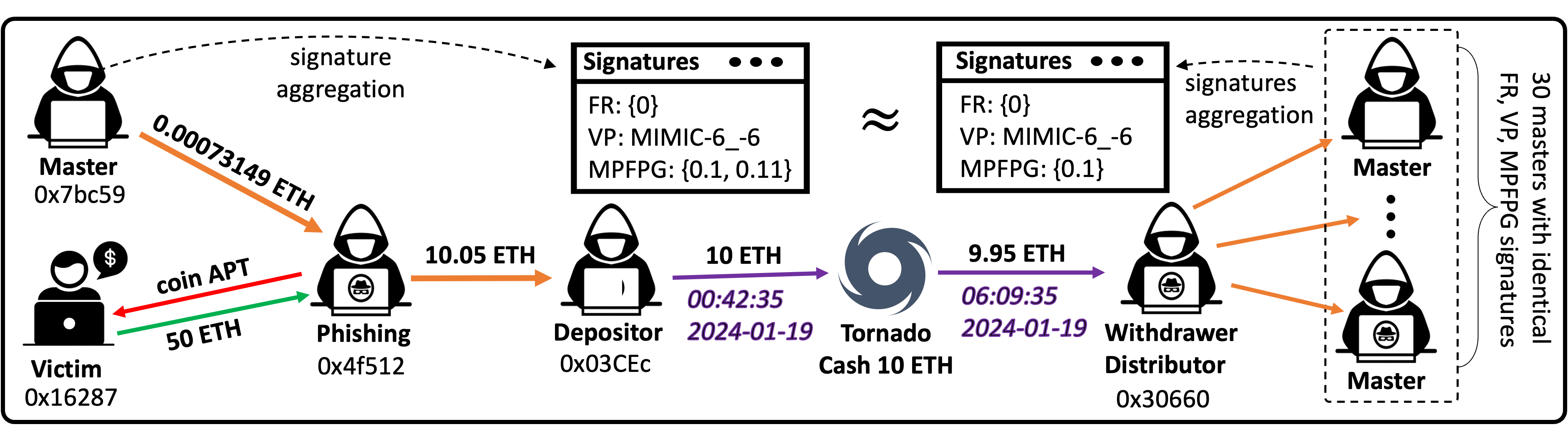}
        \caption{A potential deposit-withdrawal match for TC depositor (\protect\alinkLine{0x03CEc7583846EEEe04cDd0752A0B35bC8F793bff}) and withdrawer (\protect\alinkLine{0x306608bf57e58a1263c491ad81b1a5fa2b420cfd}) with similar FR, VP, MPFPG signatures.}
        \label{fig:TC_grouping_ex1}
    \end{figure*}

    \begin{figure*}[htb!]
        \centering
        \includegraphics[width=\textwidth]{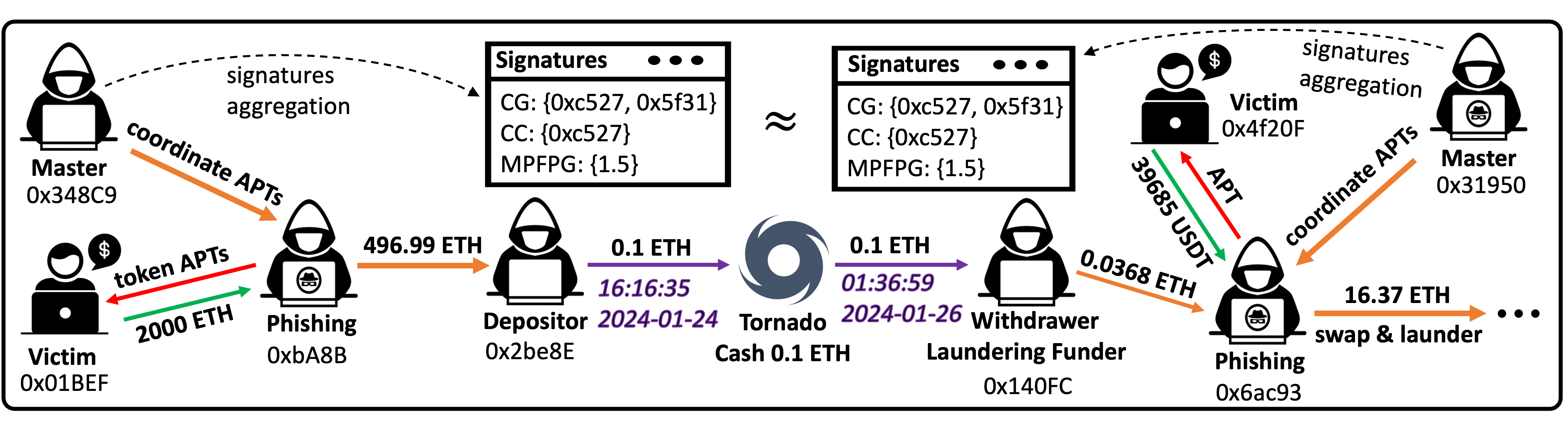}
        \caption{A pair of TC depositor (\protect\alinkLine{0x2be8E539B43b09Ef7d066C8580E12F6f1f26bEf9}) and withdrawer (\protect\alinkLine{0x140FCB5DD975078B1E394446d3F2958B7880Fd39}) that potentially belong to the same scammer/group due to similar CG, CC, and MPFPG signatures. Note that \textbf{0xc527}\ldots{ } and \textbf{0x5f31}\ldots{ } are modified addresses of the APT contract and the fake-token contracts.}
        \label{fig:TC_grouping_EX2}
    \end{figure*}

    \textbf{Tainted Paths for Depositors.} To identify the tainted depositors and the master addresses associated with each depositor, we make use of the laundering transfers between the victim addresses and the TC addresses captured in the output set $L$ of the DLF algorithm (see Section~\ref{subsec:flow_to_TC} and Appendix~\ref{app:laundering}). Using the laundering transfers from $L$, we construct a shortest laundering path from each victim transfer to each reachable depositor (there are 85 such addresses). For each depositor address, using the master dataset (see Section~\ref{subsec:master_dataset}), we find all the master addresses that coordinated or funded APTs targeting the victim before the victim transfer and involved the corresponding phishing and victim address. A \textit{tainted path} for each depositor and each related APT consists of a) the transaction initiated by the master to coordinate the APT, b) the APT-carrying transaction, c) the victim transfer, and d) the laundering path (represented by a sequence of transaction hashes) from the phishing address to TC pools (including the depositor). 
   
    \textbf{Tainted Paths for Withdrawers.} Next, we collected 16,376 unique withdrawer addresses from 35,076 TC withdrawal records (see Table~\ref{tab:TC}), and find the associated master addresses for each withdrawer as follows. We consider four types of tainted withdrawers depending on the role they play regarding APTs: Type-1 - masters (coordinate/fund APTs), Type-2 - distributors (fund masters), Type-3 - distributor funders (fund distributors), and Type-4 - laundering funders (fund successful phishing addresses after they receive the first victim transfer and before the last laundering transfer). For the withdrawers that are masters themselves, each is its own associated master. For the withdrawers that are distributors but not masters, we find their associated masters by using the distributor dataset discussed in Section~\ref{app:distributor_dataset}. We also identified withdrawers that are funders of some distributors. For such addresses, the associated masters can be found via the distributor and distributor funder datasets (see Section~\ref{subsec:master_dataset} and Appendix~\ref{app:distributor_dataset}).Finally, we include withdrawers (mostly of pools 0.1 \ETH) that are laundering funders (e.g. \alinkLine{0x388A798cA776E7781FEb7Cf9B03739703a05bd88}). Their associated masters are those that coordinated/funded the APTs leading to the laundered victim transfers.

    The tainted path for a Type-1 withdrawer (a master) contains only the transaction hash of the corresponding TC withdrawal transaction. A tainted path for a Type-2 withdrawer (a distributor) consists of the transaction hashes of the corresponding TC withdrawal transaction and the ETH transfer from the withdrawer to a master funded by the depositor. A tainted path for a Type-3 withdrawer (a distributor funder) consists of the transaction hashes of the corresponding TC withdrawal transaction, of an ETH transfer from the withdrawer to a distributor, and of an ETH transfer from the distributor to a master. Note that there can be multiple tainted paths for withdrawers of Type-2 and Type-3, whereas Type-1 withdrawers only have one tainted path. Note that the TC withdrawal transaction must happen \textit{before} the withdrawer's transfer and the transfer from distributor to master must happen \textit{after} the transfer from the distributor funder to the distributor. The tainted path for a Type-4 withdrawer (a laundering funder) consists of the transaction hash of master's transaction that funded or coordinated the APT leading to the victim transfer that requires fund laundering, the transaction hash of the transaction that carries the APT itself, the transaction hash of the victim transfer from the victim to the phishing address, the transaction hash of the TC withdrawal, and the transaction hash of the laundering funding transfer to the phishing address. Note that the laundering funding transfer must happen \textit{after} both the TC withdrawal and the victim transfer.

    \textbf{Signature Aggregation.} We consider two types of masters: coordinators and coin funders (see Section~\ref{subsec:master_types}). We retrieve the CG and the three gas signatures for coordinator masters and the FR and VP signatures for coin-funder masters (see Section~\ref{sec:signature_extraction}). For each tainted depositor and withdrawer, we assign to them the aggregated signatures from the set of the associated masters identified via the tainted paths described above. Note that during signature aggregation, we ignore useless signature values such as `OTHER' for VP signature and 21,000 for GL signature. 
    
    For VP signature, which is represented by a string, we calculate the frequencies of all string-signatures from all associated master, and assign the highest-frequency signature as the aggregated signature if its frequency exceeds a threshold $\theta=0.8$ of the sum of all frequencies. If such a high-frequency signature doesn't exist among the masters, then assign `OTHER' as the (useless) signature for the depositor/withdrawer. For signatures other than VP, which are represented by sets (after flattening - see Section~\ref{sec:signature_extraction}), we use two thresholds for aggregation. The first threshold $\gamma = 5$ and the second threshold $\theta=0.8$. THe first threshold is applied as follows: if a master's signature (as a set) has more than $\gamma$ values, then it will be excluded from aggregation. The intuition is that if the master uses many value for its signature, than their behaviour is not consistent enough to be counted. The second threshold $\theta$ is used as usual: after applying the first threshold, we calculate the frequencies of the remaining sets (signatures), and pick the sets with largest frequencies that sum up to at least $\theta$ fraction of the sum of all frequencies. We then assign the set union of such sets as the aggregated signature of the corresponding depositor/withdrawer. Otherwise, assign $\emptyset$ as the (useless) aggregated signature.

    \textbf{Deposit/Withdrawal Matching Methodology.} Scam signatures are first assigned to tainted depositors and withdrawers as before. Then, given a deposit (withdrawal), the predicted counterpart is the nearest in time withdrawal (deposit) conducted by a tainted TC user with similar scam signatures. 
    
    We discuss two examples to demonstrate the proposed methodologies. 

    \begin{example}\label{ex:TC_matching_1}
        The TC depositor \textbf{0x03CEc} (see Fig.~\ref{fig:TC_grouping_ex1}) received 10.05 \ETH from a successful phishing \textbf{0x4f512}, who got 50 \ETH from a victim thanks to a previous coin APT. This coin APT was funded by the master \textbf{0x7bc59}. Thus, the FR, VP, and MPFPG signatures of this master are assigned to the depositor. The withdrawer \textbf{0x30660}, on the other hand, after received 9.95 \ETH from TC (about 6 hours after the deposit by \textbf{0x03CEc}), funded 30 APT masters that all have identical FR, VP, and MPFPG signatures, which are then aggregated and assigned to the withdrawer. The depositor and withdrawer have identical FR and VP signatures, and also similar MPFPG signatures. Given the timing and the three highly similar signatures, the deposit and the withdrawal are potentially a match and the two addresses may be from the same scammer orgnisation.
    \end{example}

    \begin{example}\label{ex:TC_matching_2}
    Another example similar to Example~\ref{ex:TC_matching_1} is illustrated in Fig.~\ref{fig:TC_grouping_EX2}.
    \end{example}

\end{document}